\documentclass{llncs}
\usepackage[english]{babel}
\usepackage{amsmath}
\usepackage{amsfonts}
\usepackage{amssymb}
\usepackage{booktabs} 
\usepackage[a4paper, total={6in, 8in}]{geometry}
\usepackage{yhmath}
\usepackage{float}
\usepackage{authblk}
\usepackage[table]{xcolor}
\usepackage{hyperref}
\hypersetup{colorlinks,
			citecolor=Tapestry Blue,
			urlcolor=Prussian Blue,
			linkcolor=Raspberry Red,
                anchorcolor=Charcoal Gray}
\usepackage{comment}
\usepackage{forest} 
\usepackage{multirow}

\usepackage{soul}
\usepackage[operators,sets]{cryptocode}
\usepackage{msc} 

\usepackage{dblfloatfix}

\definecolor{Charcoal Gray}{HTML}{1C1C1C}
\definecolor{Prussian Blue}{HTML}{113285}
\definecolor{Tapestry Blue}{HTML}{0C4842}
\definecolor{Raspberry Red}{HTML}{8E354A}
\definecolor{Midnight Blue}{HTML}{0B1013}

\newcommand{\CL}{CL}
\newcommand{\BBS}{BBS+}
\newcommand{\bbs}{BBS}
\newcommand{\PS}{PS}
\newcommand{\sphincs}{SPHINCS\textsuperscript{+}}
\newcommand{\cm}{cm}
\newcommand{\ARI}{A}
\newcommand{\salt}{S}
\newcommand{\DA}{DA}
\newcommand{\DS}{DS}

\newcommand{\SDS}{CMT}
\newcommand{\comList}{cmtList}
\newcommand{\merTree}{merTree}
\newcommand{\SDSig}{SDSig}
\newcommand{\sig}{\sigma}
\newcommand{\TT}[1]{\mathtt{#1}}
\newcommand{\HH}{\mathcal{H}}
\newcommand{\pp}{\TT{pp}}
\newcommand{\setUp}{setUp}
\newcommand{\keyGen}{keyGen}
\newcommand{\sign}{genSig}
\newcommand{\ver}{verSig}
\newcommand{\genIssProof}{genIssuerProof}
\newcommand{\verIssProof}{verIssuerProof}
\newcommand{\genHolderProof}{genHolderProof}
\newcommand{\verPresProof}{verPresentProof}

\newif\ifFIXESON
\FIXESONtrue 
\ifFIXESON
\newcommand{\fixn}[2]{\fixfootnote{\textbf{#1:} #2}}
\usepackage[draft,inline,nomargin,index]{fixme}
\else 
\newcommand{\fixn}[2]{}
\usepackage[final]{fixme}
\fi
\fxsetup{theme=color,mode=multiuser}
\definecolor{jade}{HTML}{00A86B}
\definecolor{tan}{HTML}{D2B48C}
\FXRegisterAuthor{gs}{envgs}{\color{jade}GS}  
\FXRegisterAuthor{at}{envat}{\color{Midnight Blue}AT}   
\FXRegisterAuthor{af}{envaf}{\color{blue}AF}   
\FXRegisterAuthor{as}{envsm}{\color{purple}AS}  
\FXRegisterAuthor{ms}{envms}{\color{red}MS}   
\fxsetup{layout={footnote}}
\fxusetheme{colorsig}

\newcommand{\high}{$+$}

\newcommand{\highhh}{$+++$}

\newcommand{\meh}{$\pm$}

\newcommand{\low}{$-$}
\newcommand{\loww}{$--$}

\definecolor{cb_india}{HTML}{e5af53}
\definecolor{cb_cyan}{HTML}{0173b2}
\definecolor{cb_aquamarine}{HTML}{029e73}
\definecolor{cb_lilac}{HTML}{cc78bc}
\definecolor{cb_caramel}{HTML}{df8f50}
\definecolor{cb_dandelion}{HTML}{ece133}

\newcommand{\cellhigh}{\cellcolor{cb_aquamarine!20}\high}

\newcommand{\cellhighhh}{\cellcolor{cb_aquamarine!60}\highhh}

\newcommand{\cellmeh}{\cellcolor{cb_dandelion!20}\meh}

\newcommand{\celllow}{\cellcolor{cb_caramel!20}\low}
\newcommand{\cellloww}{\cellcolor{cb_caramel!40}\loww}

\title{On Cryptographic Mechanisms for the Selective Disclosure of Verifiable Credentials}
\author{Andrea Flamini\inst{1}\orcidID{0000-0002-3872-7251} \and Giada Sciarretta\inst{2}\orcidID{0000-0001-7567-4526} \and
Mario Scuro\inst{1}\orcidID{0000-0003-2410-3760} \and Amir Sharif\inst{2}\orcidID{0000-0001-6290-3588} \and Alessandro Tomasi\inst{2}\orcidID{0000-0002-3518-9400} \and Silvio Ranise\inst{1,2}\orcidID{0000-0001-7269-9285}}

\institute{Department of Mathematics, University of Trento, Trento, Italy 
\and
Center for Cybersecurity, Fondazione Bruno Kessler, Trento, Italy }

\begin{document}
\maketitle

\begin{abstract}
Verifiable credentials are a digital analogue of physical credentials. Their authenticity and integrity are protected by means of cryptographic techniques, and they can be presented to verifiers to reveal attributes  or even predicates about the attributes included in the credential.
One way to preserve privacy during presentation consists in selectively disclosing the attributes in a credential. In this paper we present the most widespread cryptographic mechanisms used to enable selective disclosure of attributes identifying two categories: the ones based on hiding commitments - e.g., m\textsc{dl} ISO/IEC 18013-5 - and the ones based on non-interactive zero-knowledge proofs - e.g., BBS signatures. We also include a description of the cryptographic primitives used to design such cryptographic mechanisms. 

We describe the design of the cryptographic mechanisms and compare them by performing an analysis on their standard maturity in terms of standardization, cryptographic agility and quantum safety, then we compare the features that they support with main focus on the unlinkability of presentations, the ability to create predicate proofs and support for threshold credential issuance. 

Finally we perform an experimental evaluation based on the Rust open source implementations that we have considered most relevant. In particular we evaluate the size of credentials and presentations built using different cryptographic mechanisms and the time needed to generate and verify them. We also highlight some trade-offs that must be considered in the instantiation of the cryptographic mechanisms.


\end{abstract}

\section{Introduction}
\label{intro}

As more services move online, increasing importance is given to an individual's digital identity as the foundation for secure and trusted online interactions, related to e-government and e-commerce.

A new paradigm for identity management based on digital identity wallets is emerging to empower data subjects to selectively disclose the user attributes within what is called verifiable credentials in a privacy-preserving and secure way. A verifiable Credential is a digital attestation or evidence of particular information about an individual that is intended to be cryptographically secure, and tamper-proof.
The most prominent example of the aforementioned paradigm is the revised regulation eIDAS 2~\cite{european2021regulation},  
proposing a European Digital Identity (EUDI) wallet that can be used by the user to securely store the issued verifiable credentials and aims to improve cross-border interoperability. The privacy-enhancing aims of the EUDI wallet include offering data subjects the means to control who has access to which of their personally identifiable information, and making it possible to selectively disclose only some of the attributes in their verifiable credentials to trusted parties. 
When a service provider requests too many subject claims, it may 
dissuade users from utilizing the service. Furthermore, extensive data collection 
increases the risk of data breaches or misuse, and does not follow 
data minimization and privacy by design principles under the GDPR~\cite{GDPR}, a basic part of data protection.

In the design of their protocols and implementations, service providers must consider trade-offs between simplicity vs sophistication of protocol, implementation, and deployment issues including resource constraints.



\paragraph{Scenario.}

To exemplify selective disclosure, we consider the following simplified scenario: a subject wishes to purchase alcohol and to prove that (s)he is over the legal age limit in the jurisdiction, e.g., 18, without fully disclosing her entire mobile driving license (m\textsc{dl}).

In this example, the agency in charge of issuing m\textsc{dl} (Issuer) verifies the m\textsc{dl} Subject's age during the issuance process and includes it as an attribute in the m\textsc{dl}. The data Subject holding the m\textsc{dl} can select to disclose the single m\textsc{dl} attribute ``age'' to the liquor store employee (Verifier).
The Verifier can check that the Subject is of age to buy alcohol without learning any other personal information. 

This enhances privacy for the Subject while enabling the Verifier to check their age while complying with the data minimization principle. 

\paragraph{Contributions.}
eIDAS 2 states that EUDI wallets ``should technically enable the selective disclosure of attributes within verifiable credentials'', and amendments to the proposal add ``where attestation of attributes does not require the identification of the user, zero knowledge attestation shall be performed''~\cite{eIDAS2_amendments}. The EUDI Wallet Architecture and Reference Framework (ARF)~\cite{EUDI_ARF}, intended to provide more concrete technical guidelines and tools, states that ``attestation MUST enable Selective Disclosure of attributes by using Selective Disclosure for JSON Web Tokens (SD-JWT) and Mobile Security Object (ISO/IEC 18013-5) scheme''.

Both schemes cited in the ARF are based on hiding commitment mechanisms - generating a commitment to a value while keeping it hidden, with the ability to reveal the committed value later~\cite{jwp4mertree}. The ARF does not currently cover zero knowledge proofs (ZKP) - e.g., repeatedly proving knowledge of a value without ever having to reveal it~\cite{IdeMix,anoncreds_1_spec_Ursa,anoncreds_2_spec_Ursa}. Given the complexity and range of available options, it is non-trivial to assess the pros and cons of each option. In order to facilitate an informed choice, we provide cryptographic building blocks for credentials with selective disclosure capability based on hiding commitments and ZKP. In short, we extend our work in~\cite{secrypt23} and make the following main contributions:
\begin{itemize}
    \item We summarize six cryptographic mechanisms (\texttt{\cm}) for selective disclosure based on hiding commitment and ZKP, providing more detail than~\cite{secrypt23} and two new \texttt{\cm}, BBS (Section \ref{subsec:BBS}) and PS (Section \ref{subsec:PS}) signatures.

    \item We provide the structure of Verifiable Credentials and Presentations for the \texttt{\cm}, together with the operation of entities that must be performed for their creation (issuing) and consumption (presentation).
    
    \item We compare the \texttt{\cm} w.r.t.\ several features to assist in selecting the most appropriate for the use case of interest.

    Our analysis has been expanded over~\cite{secrypt23} by considering the following features:  quantum safety of cryptographic algorithms, support for threshold credential issuance, and an analysis of trade-offs that lead to interesting implementation choices in the solutions we examined. While the first one has been considered to evaluate the maturity of cryptographic mechanisms w.r.t. quantum resistance, the rest have been considered to evaluate for each cryptographic mechanism how they support features that are relevant to the design and implementation of practical privacy preserving Verifiable Credentials. 

\end{itemize}

\paragraph{Outline.}
Section \ref{sec:selectiveDisc} introduces the verifiable credential ecosystem, the formats of verifiable credentials that support the selective disclosure of attributes, and their lifecycle. Section \ref{sec:cryptoBuiliding} introduces the cryptographic primitives used to implement the cryptographic mechanisms described in Sections \ref{sec:hiding-commitment} and \ref{sec:sd-signatures}. In Section \ref{sec:secAnalys}, we analyse the mechanisms and we discuss how they support some privacy-enhancing features. In Section \ref{subsec:evaluation} we perform an experimental evaluation of the mechanisms described.  We summarize the main results and discuss future work in Section \ref{sec:conclusion}.

\section{{Verifiable Credentials and Selective Disclosure}}
\label{sec:selectiveDisc}
Following the Verifiable Credential data model~\cite{VC_data_model}, a credential can be defined as ``a set of one or more claims [assertions about a Subject] made by an Issuer'', and a Verifiable Credential (VC) as ``a tamper-evident credential that has authorship that can be cryptographically verified''. We consider the following entities and quote the descriptions from \cite{OpenID4VC}:

\begin{description}
    \item[Issuer:] ``a role an entity can perform by asserting claims about one or more subjects, creating a VC from these claims, and transmitting the VC to a holder''.
    \item[Holder:] ``a role an entity might perform by possessing one or more VCs and generating presentations from them''.
    \item[Subject:] ``the entity about which claims are made''.
    \item[Verifier:] ``a role an entity performs by receiving one or more VCs, optionally inside a verifiable presentation'' and verifies it ``to make a decision regarding providing a service to the Subject".
\end{description}

We describe the general structure of VCs and Verifiable Presentations (VPs) regardless of the cryptographic mechanism used.

A VC is composed of three sections: an \emph{Issuer protected header}, containing general information about the credential, for instance the Issuer, the Subject and the credential type, an \emph{Issuer payload} containing information about the credential attributes, and an \emph{Issuer proof} which contains the cryptographic material which attests the authenticity of the credential (see Table \ref{tab:cred}).

A VP is composed of three sections (see Table \ref{tab:pres}): a \emph{presentation protected header} with general information about the credential; a \emph{presentation payload} with information related to the disclosed attributes; and a \emph{presentation proof} with the cryptographic material that allows the Verifier to check the authenticity of the presentation.

The structure of the VC and VP we adopt is consistent, albeit simplified to focus on selective disclosure, with the structure of JSON Web Proof (JWP) \cite{JWP}, a proposal
to standardize a JSON container which aims to describe the structure of VCs to allow the selective disclosure of attributes.

In a preliminary \emph{set-up phase}, the Issuer must generate its private-public key pair $(\TT{sk_{Iss}},\TT{pk_{Iss}})$ using the key generation function of the digital signature scheme used to sign the VCs, $\TT{\keyGen}()$. In the \emph{issuing phase}, the Issuer generates an Issuer proof with the function $\TT{\genIssProof}(-)$. The Holder, upon reception of the VC created by the Issuer, verifies its validity computing the function $\TT{\verIssProof}(-)$.

In the \emph{presentation phase} the Holder can create a VP specifying the attributes it wants to disclose. In particular, the Holder creates the VP containing the \emph{Holder-generated proof} by computing the function\\ $\TT{\genHolderProof}(-)$. The Verifier, upon reception of the VP computes the function $\TT{\verPresProof}(-)$ to verify it and possibly accept the Holder's claims.

 \begin{table*}[!b]
  \centering
 {
 \small
 \caption{A simplified representation of VC which allows for the selective disclosure of attributes.}
 \label{tab:cred}
 \begin{tabular}{||l|l|l||} 
 \hline
 VC &  \textbf{Hiding-commitment}  & \textbf{Selective disclosure signature}  \\
 \hline
 \hline
    \textbf{Issuer Protected Header} &  Cryptographic mechanism: $\TT{\cm}$ & Cryptographic mechanism: $\TT{\cm}$ \\
    & Issuer public key: $\TT{pk_{Iss}}$  &  Issuer public key: $\TT{pk_{Iss}}$   \\
 \hline
    \textbf{Issuer Payloads} & Attributes and salts:  &  Attributes:\\
     & \hspace{10pt}$\TT{\ARI}=(a_1,\dots,a_m)$ &   \hspace{10pt}$\TT{\ARI}=(a_1,\dots,a_m)$\\
     & \hspace{10pt}$\TT{\salt}=(s_1,\dots,s_m)$& \\
 \hline
    \textbf{Issuer Proof} & Signed commitment: & Selective disclosure signature:\\
                          
                          & \hspace{10pt}$\TT{\genIssProof(sk_{Iss},\ARI,\salt)=}$  & \hspace{10pt}$\TT{\genIssProof(sk_{Iss},\ARI)=}$\\
                          & \hspace{10pt}$\TT{=(\SDS,\sig=\sign{(sk_{Iss},\SDS))}}$ &
                          \hspace{10pt}$\TT{=\sig=\sign(sk_{Iss},\ARI})$ \\
 \hline
\end{tabular}
}
 \end{table*} 

  \begin{table*}[!b]
    \centering
     \small
 {
 \caption{The general structure of a VP derived from a VC as in Table \ref{tab:cred}.}
 \label{tab:pres}
 \begin{tabular}{||l|l|l||} 
 \hline
 VP&  \textbf{Hiding-commitment}  & \textbf{Selective disclosure signature}  \\
 \hline
 \hline
    \multirow{2}{3cm}{\textbf{Presentation Protected Header}}&  Cryptographic mechanism: $\TT{\cm}$ & Cryptographic mechanism: $\TT{\cm}$ \\
    & Issuer public key: $\TT{pk_{Iss}}$  &  Issuer public key: $\TT{pk_{Iss}}$   \\
 \hline
    \multirow{2}{3cm}{\textbf{Presentation Payloads}} &  Disclosed attributes and salts:  &   Disclosed attributes: \\
     & \hspace{10pt}$\TT{\DA}=(a_{i_1},\dots,a_{i_d})\subset \TT{\ARI}$ &   \hspace{10pt}$\TT{\DA}=(a_{i_1},\dots,a_{i_d})\subset \TT{\ARI}$\\
     & \hspace{10pt}$\TT{\DS}=(s_{i_1},\dots,s_{i_d})\subset \TT{\salt}$ &   \\
 \hline
   \multirow{2}{3cm}{\textbf{Presentation Proof}} & Signed commitment: & \\
                          & \hspace{10pt}$\left( \TT{\SDS}, \TT{\sig} \right)$  & \\
                          &  Holder-generated Proof: &   Holder-generated proof:  \\
                         &   \hspace{10pt}$P=\TT{\genHolderProof(\DA,\DS,\ARI,\salt)}$&  \hspace{10pt}$P=\TT{\genHolderProof(pk_{Iss},\DA,\ARI,\sig)}$    \\
 \hline
\end{tabular}
}
 \end{table*}

\subsection{Taxonomy of Cryptographic Techniques for VC Selective Disclosure}
There are several methods that allow VCs to support selective disclosure. \cite{VC_imp_guide}, identifies the following categories: \emph{atomic credentials}, \emph{hashed values} and \emph{selective disclosure signatures} (which in literature are also referred to as \emph{anonymous credentials}\cite{IdeMix,camenisch2016anonymous,camenisch2002signature}). Atomic credentials contain only a single attribute, therefore the Issuer may provide a set of atomic credentials, then the Holder presents to a Verifier only those that it wants to show. Atomic credentials are unwieldy to manage, particularly to guarantee that a presentation contains a collection of atomic credentials that is valid as a whole, but do not introduce or require substantially different cryptographic techniques than the other two mechanisms; therefore, we do not discuss them further. Instead we focus on the other two categories of mechanisms:
\begin{description}
    \item \textbf{Hashed values} allow an Issuer to issue a single VC containing multiple claims.
    Each claim is hidden and committed to using hash functions, then the commitment is signed by the Issuer. Examples include hash lists (Section \ref{subsubsec:comlist}) and Merkle trees (Section \ref{subsub:merTree}).
    \item \textbf{Selective disclosure signatures} are signatures schemes that natively support selective disclosure of VC claims by using non-interactive zero knowledge proofs. Examples are \emph{CL} (Section \ref{subsec:CL}), \emph{BBS} (Section \ref{subsec:BBS}), \emph{BBS+} (Section \ref{subsec:BBS+}) and \emph{PS} (Section \ref{subsec:PS}) signatures.
    
\end{description}

We provide noteworthy examples of cryptographic mechanisms based on hashed values, considered as an instance of hiding commitments, which are adopted in the standardized mobile Driving License \cite{ISO_18013-5} or discussed in \cite{jwp4mertree} (Section \ref{sec:hiding-commitment}). We also present examples of the most relevant selective disclosure signatures adopted in \cite{IdeMix,anoncreds_1_spec_Ursa,anoncreds_2_spec_Ursa}(Section \ref{sec:sd-signatures}).

In Table \ref{tab:acr} we report all the acronyms, functions and variables used in the paper.

\begin{table*}[!b]
  \centering
 {
 \small
 \caption{List of acronyms, functions and variables.}
 \label{tab:acr}
 \begin{tabular}{||l|l||} 
 \hline
 $\TT{\keyGen(-)}$ & digital signature key generation algorithm\\
 \hline
 $\TT{\sign(-)}$ & signature generation algorithm\\
 \hline
 $\TT{\ver(-)}$ & signature verification algorithm\\
 \hline
 $\TT{\genIssProof(-)}$ & Issuer proof generation algorithm\\
 \hline
 $\TT{\verIssProof(-)}$ &  Issuer proof verification algorithm\\
 \hline
 $\TT{\genHolderProof(-)}$ & Holder-generated proof generation algorithm\\
 \hline
 $\TT{\verPresProof(-)}$ & Holder-generated proof verification algorithm\\
 \hline
 VC & verifiable credential\\
 \hline
 VP & verifiable presentation\\
 \hline
 $\TT{\cm}$ & cryptographic mechanism\\
 \hline
 HVZK & honest verifier zero-knowledge\\
 \hline
 NIZKP & non-interactive zero-knowledge proof\\
 \hline
 $\TT{\ARI}$ & list of attributes included in the VC\\
 \hline
 $\TT{\salt}$ & list of salts included in the VC based on hiding commitments\\
 \hline
 $\TT{\DA}$ & disclosed attributes included in the VP\\
 \hline
 $\TT{\DS}$ & disclosed salts included in the VP\\
 \hline
 $\TT{\SDS}$ & commitment included in VC and VP based on hiding commitment\\
 \hline
 $\TT{\comList}$ & list of hash and salt cryptographic mechanism\\
 \hline
 $\TT{\merTree}$ & Merkle tree cryptographic mechanism\\
 \hline
 $\TT{\SDSig}$ & selective disclosure signature\\
 \hline
 $\sig$ & output of any digital signature algorithm\\
 \hline
 $\HH(-)$ & cryptographic hash function\\
 \hline
\end{tabular}
}
 \end{table*}

\section{{Background on Cryptographic Building Blocks}}
\label{sec:cryptoBuiliding}
We provide the main cryptographic notions that are useful to understand the approaches for the creation of VCs supporting selective disclosure of attributes: digital signatures (Section \ref{subsec:digsig}), hashing and salting for the creation of hiding commitments (Section \ref{subsec:hiding commitment}), and NIZKP (Section \ref{subsec:NIZKP}) to prove statements about undisclosed attributes in selective disclosure signatures. We also highlight the threat models that are useful to understand the security properties satisfied by each cryptographic primitives.

\subsection{Digital signatures}
\label{subsec:digsig}

In cryptographic mechanisms based on hiding commitment or selective disclosure signature, the essential cryptographic tool used to prove the authenticity of a VC, and the validity of the derived VPs, are the digital signature algorithms used by the Issuer to sign the VC when issuing it to the Holder.

Digital signature schemes are defined by the algorithms \texttt{\setUp($\lambda$)} to generate public parameters \texttt{pp} given a security level $\lambda$, \texttt{\keyGen(pp)} to generate the public-private key pair $(\TT{pk,sk})$, $\TT{\sign(sk},m)$ to sign a message $m$, and $\TT{\ver(pk},m,\TT{\sig})$ to verify the signature $\sig$. 

While the digital signature schemes we use in hiding commitment-based cryptographic mechanisms (Section \ref{sec:hiding-commitment}) may be any standardized digital signature algorithm, those used in selective disclosure signature-based cryptographic mechanisms (Section \ref{sec:sd-signatures}) are a special class of signatures designed to support ZKP, and may require more structured inputs, e.g., ordered lists of messages. We use the same notation for brevity, but we stress that they support different features that additionally require the generation of public parameters that are specific to individual attributes in a credential. Depending on the algorithm and the trust model, these elements may be included in the list of public parameters, or be part of the public key. We summarize these distinctions in Section \ref{subsec:set-up}.

\paragraph{Threat model.} 
According to \cite{katz2007introduction}, Section 13.2, given a public key $pk$ generated by a signer $S$, a \emph{forgery} is a valid signature $\sigma$ (verifiable using $pk$) of a message $m$ not previously signed by $S$, and a signature scheme is secure (or unforgeable) if an adversary is not able to produce a forgery. A signature scheme is said to be \emph{unforgeable under a chosen message attack} if it is secure against an adversary with the power to ask $S$ to provide the signature of many messages of its choice before producing a forgery. All the digital signatures that we describe in this paper are proven unforgeable under a chosen message attack. 

To prove a signature secure, or more precisely unforgeable under a chosen message attack, it is necessary to prove that if an attacker with the above capability - querying $S$ and creating a valid forgery - did exist, it could be used as a subroutine of an attacker who can win a different experiment $\mathcal{E}$ that is believed infeasible to win. In this case we will say that ``experiment $\mathcal{E}$ is hard to win'' is the assumption under which the digital signature is secure under a chosen message attack. As long as the assumption holds, no attacker should be able to forge a signature.

The digital signatures described in Section \ref{sec:sd-signatures} are based on different assumptions. 

In the use case of our interest, breaking the unforgeability of the signature would give an attacker the ability to create new VCs that could be verified using the Issuer's public key. 

\subsection{Hiding commitments}
\label{subsec:hiding commitment}
Informally, a \emph{commitment scheme} allows a party to commit to a value $v$ by sending a commitment, and then to reveal $v$ by \emph{opening} the commitment at a later point in time. The commitment scheme must satisfy the \emph{binding} property, which is: a commitment to a value $v$ can not be opened to a value $v'\ne v$. A \emph{hiding commitment scheme} must satisfy also the \emph{hiding} property: from the commitment it must not be feasible to retrieve the committed value $v$.

Since our goal is to describe the design of VCs that allow the selective disclosure of attributes, we are interested in hiding commitment schemes that take as input an ordered list of values such that the opening algorithm can be performed on specific positions of the list. The hiding and the binding properties must hold on the ordered list of commitments. They are adapted in the following way: it must be infeasible to retrieve the values in the positions of the list that do not get opened, and it must be infeasible to open a position of the list to a different value than the one used to create the commitment, a property often referred to as position binding \cite{catalano2013vector}.

The VC created using hiding commitment based cryptographic mechanisms instruct the Issuer to create an hiding commitment to the attributes it want to include in the VC, then to sign it. Signing the commitment, the Issuer implicitly also signs the attribute used to create it. At a later point in time, when the Holder wants to present the VC, it shows the signature of the commitment to the Verifier, and thanks to the hiding property of the hiding commitment, this does not reveal any information about the attributes used to create it. Therefore, the attribute behind the commitment can be kept hidden if the Holder does not want to disclose it to a Verifier. On the other hand, if the Holder wants to reveal an attribute, it can open the signed commitment and show that the Issuer has certified it. Note that the Holder can not open a commitment to a different message from the one used to create it, thanks to the position binding property of the commitment scheme.

\subsubsection{Hash and Salt Technique}
A widely adopted approach for the creation of hiding commitments is based on cryptographic hash functions. Cryptographic hash functions satisfy very important security properties such as the \emph{preimage resistance} property that informally states that, given a digest $y$ it is infeasible to find an input $x$ whose digest $\HH(x)=y$, and the \emph{collision resistance} property which states that it is infeasible to find $x,y$ such that $\HH(x)=\HH(y)$.

The commitment scheme based on cryptographic hash functions is defined as follows: the commitment creation algorithm takes as input a value ${v}$ to be committed to, and outputs ${\HH(v||s)}$, where ${s}$ is chosen uniformly at random and is referred to as the salt of the commitment, and $v||s$ is the concatenation of the bytes strings $v$ and $s$.

\paragraph{Threat model.}
Similarly to digital signature schemes, the security of commitment schemes are defined using experiments that capture the security properties that a hiding commitment scheme must satisfy, namely the hiding property and the binding property.

According to \cite{katz2007introduction}, Section 6.6.5, the experiment used to prove that a commitment is hiding is the following: the attacker chooses two messages $m_1$ and $m_2$ and sends it to the challenger. The challenger chooses at random one of the two messages, creates a commitment to it and sends it to the attacker. The attacker must decide which of the two messages has been used to create the commitment. 
For what concerns the binding property, the experiment used to prove that a commitment is binding requires the attacker to give to the challenger a single commitment together with two distinct messages and associated opening material that allow to open the commitment to each of the two messages.

The hiding property of the commitment based on the hash and salt technique is derived from the preimage resistance property of the underlying hash function $\HH$ and the binding property of the commitment is derived from the collision resistance property of $\HH$.

If an attacker were capable of breaking the hiding property, it would be able to learn information about the attributes that the Holder wants to keep hidden during a verifiable presentation of a VC as the ones described in Section \ref{sec:hiding-commitment}. If an attacker could break the binding property it would be able to open a commitment to two different values, therefore it would be able to present, in distinct VP, different values for the same attribute of the same VC as the ones described in Section \ref{sec:hiding-commitment}. 

\subsection{Non-Interactive Zero-Knowledge Proofs}
\label{subsec:NIZKP}

Non-interactive zero-knowledge proofs (NIZKP) for a relation $\mathcal{R}\subset W\times Y$ where $W$ is the set of witnesses and $Y$ the set of statements, allow an actor, called prover, to convince another actor, called verifier, that it knows a witness $w$ for a statement $y$ without revealing anything else to the verifier. The protocol is non-interactive, meaning that the prover generates a proof $\pi$ and the verifier checks that $\pi$ is valid without requiring additional interactions between prover and verifier.

\paragraph{Signature Proof of Knowledge (SPK).}
For the sake of brevity, along the paper we will adopt the notation introduced in \cite{camenisch1997efficient} and we write \[\pi \in SPK\{(w_1,\dots,w_n):y=\prod_{i=1}^ng_i^{w_i}\}\] to represent a NIZKP of knowledge of a witness $(w_1,\dots,w_n)\in W$ for the statement $y\in Y$ such that $y=\prod_{i=1}^ng_i^{w_i}$. The NIZKP used refers to the relation \[\mathcal{R}=\{((w_1,\dots,w_n),y)|y=\prod_{i=1}^ng_i^{w_i}\}\subset W\times Y\] and is referred to as NIZKP for linear relations which is a main building block for the cryptographic mechanisms presented in Section \ref{sec:sd-signatures}. In \ref{sigmaFS}, Figure \ref{fig:sigma_fiat_shamir}, we provide a description this algorithm.

In Section \ref{sec:sd-signatures} we use NIZKP in combination with a special class of digital signatures, referred to as selective disclosure signatures. In particular, the Issuers create VCs by signing the attributes using this kind of digital signature, and issue it by giving the signature and the attributes to the Holder. Later the Holder can prove knowledge of such signature on the set of attributes it wants to disclose to a Verifier in zero knowledge using the NIZKP associated to each selective disclosure signature. The Verifier will only learn that the Holder knows a signature made by the Issuer over the disclosed attributes. It can not learn any information about the signature and about the hidden attributes. 
We will describe four NIZKPs based on the NIZKP for linear relations: the first, in Section \ref{subsec:CL}, is based on a variant of the sigma protocol for linear relation adapted to work having as set of statements $Y$ a group of unknown order, whereas in Section \ref{subsec:BBS}, Section \ref{subsec:BBS+} and Section \ref{subsec:PS} the set of statement $Y$ will be a group of prime order $p$. 

\paragraph{Threat model.}

According to \cite{boneh2020graduate}(Attack Game 20.3), the threat model considered for NIZKPs is the following. A challenger offers one of two games to an attacker, a ``real world'' game and a ``simulated world'' game, without revealing which one is being offered. The attacker must try to distinguish which of the two games it is playing, judging by the challenger's responses. The attacker sends the challenger a pair $(w,y)\in \mathcal{R}$, and asks for a proof about it. If the real world game is being played, the challenger creates a proof $\pi$ as prescribed by the NIZKP using a real random oracle; if the simulated world game is being played, the challenger (also called simulator) creates a simulated proof without using the knowledge of $w$, as if it does not know $w$, by simulating also the random oracle\footnote{This extra power that we give to the simulator is crucial: the protocol must be a proof of knowledge of the witness, i.e., a protocol whose output is proof that can be generated only by someone who knows the witness.} by programming it according to the queries it receives from the attacker as in \cite{abdalla2002identification}, Lemma 3.5.

The protocol is a NIZKP if every attacker has a negligible advantage in distinguishing whether it is performing the real world experiment or the simulated world experiment.
A more detailed discussion on the way the NIZKP are built is reported in \ref{sigmaFS} and in \cite{boneh2020graduate} (Section 20.3.5).

An attacker who can distinguish the real world from the simulated world might be able to learn some information related to the witness known by the prover. In the case of selective disclosure signatures for VCs as in Section~\ref{sec:sd-signatures}, this might imply the ability for a Verifier to gain information about the signature of the VC used by the Holder, or about the hidden attributes.

\section{{Hiding-commitment Mechanisms}}
\label{sec:hiding-commitment}

 Instances of hiding commitment mechanisms can be obtained by using lists of hash-based hiding commitments ($\TT{\comList}$, see Section \ref{subsubsec:comlist}), or Merkle Trees ($\TT{merTree}$, see Section \ref{subsub:merTree}), as suggested in \cite{jwp4mertree}.

The Issuer commits to a set of attributes, then digitally signs the commitment.
The properties of hiding commitments allow the Issuer of a credential to sign the commitments, then a Holder, who knows the attribute values of a credential, can open only some of the committed values proving to a Verifier the truthfulness of its claims. The security of the schemes we describe below resides on the security of the digital signature used, as discussed in Section \ref{subsec:digsig}, and on the security of the hiding commitment schemes as discussed in Section \ref{subsec:hiding commitment}.

\paragraph{Operations in the Issuing Phase.} 

The Issuer can create a VC with the structure of Table \ref{tab:cred} and issues it to the Holder. The VC is composed of the three parts already mentioned:
\begin{itemize}
    \item the \emph{Issuer protected header} containing the cryptographic mechanism identifier $\TT{\cm}$, - specifying primitives such as the chosen digital signature algorithm and cryptographic hash function - and the Issuer public key $\TT{pk_{Iss}}$;
    
    \item the \emph{Issuer payload} containing a list of attributes $\TT{\ARI}=(a_1,\dots,a_m)$ certified by the Issuer who created the credential, together with a list of random salts, one for each attribute $\TT{\salt}=(s_1,\dots,s_m)$;
    
    \item the \emph{Issuer Proof} containing the digital signature of the commitment $\TT{\SDS}$ to the attributes $\TT{\ARI}$, constructed according to the chosen cryptographic mechanism and the list of attributes and salts, signed by the Issuer, obtaining ${\sig}=\TT{\sign(sk_{Iss},CMT)}$. These operations are performed executing the function $\TT{\genIssProof(sk_{Iss},\ARI,\salt)}$.
\end{itemize}
Note that the choice of the digital signature scheme adopted by the Issuer to sign the $\TT{\SDS}$ is not restricted to a specific primitive.

The Holder can verify the VC's validity by computing the function $\TT{\verIssProof(VC)}$, which consists in verifying that the commitment $\TT{\SDS}$ is actually a commitment to the elements in $\TT{\ARI}$ and $\TT{\salt}$, and verifying the Issuer's digital signature.

\paragraph{Operations in the Presentation Phase.}
The Holder creates a VP to convince the Verifier that the attributes revealed are included in a credential issued by a trusted Issuer.

A VP in this context has the structure described in Table \ref{tab:pres}. It is composed by:
\begin{itemize}
    \item a \emph{presentation protected header} containing the name of the cryptographic mechanism $\TT{\cm}$ adopted in the creation of the underlying credential and the Issuer public key;
    
    \item the \emph{presentation payloads}, containing a subset $\TT{\DA\subset \ARI}$ of attributes $(a_{i_1},\dots, a_{i_d})$ that the Holder wants to disclose together with $\TT{\DS\subset\salt}$, the list of associated salts $(s_{i_1},\dots,s_{i_d})$;
    
    \item a \emph{presentation proof} generated by the Holder including the commitment $\TT{\SDS}$ and its signature ${\sig}$ created by the Issuer associated to $\TT{pk_{Iss}}$ and the Holder-generated proof obtained computing the function $\TT{\genHolderProof(\DA,\DS,\ARI,\salt)}$.
\end{itemize}

The Verifier verifies a VP received from the Holder by computing the function $\TT{\verPresProof(VP)}$, which consists in $(i)$ verifying the signature of the $\TT{\SDS}$ created by the Issuer, and $(ii)$ verifying the proof that the disclosed attributes in $\TT{\DA}$ are a subset of the attributes committed to in $\TT{\SDS}$.\\

Once 
the commitment opening algorithm for the pairs $(a_i,s_i)$ in $\TT{\DA\times\DS}\subset\TT{\ARI\times\salt}$ is defined, the functions $\TT{\genHolderProof(\DA,\DS,\ARI,\salt)}$ and $\TT{\verPresProof(VP)}$ are well defined.

\subsection{Commitment List Mechanism}
\label{subsubsec:comlist}

In the $\TT{\comList}$ mechanism, credentials contain ordered lists of attribute-salt pairs; for each pair, the issuer creates a hiding commitment, then signs the list of commitments.

In $\TT{\genIssProof(sk_{Iss},\ARI,\salt)}$, the Issuer generates a random salt $s_i$ for each attribute ${a_i}$ and computes the commitment list entries $L_i=\HH(a_i||s_i)$. Finally, $\TT{\SDS}=\left[L_i\right]_{i=1}^{\#\ARI}$ is signed by the Issuer to create the Issuer proof.

Since the payload of a Holder-generated VP (Table~\ref{tab:pres}, column 2) contains all the information needed to open the commitments to the disclosed attributes, the Presentation Proof only contains the signed commitment i.e. \texttt{\genHolderProof} is the null function.

In $\TT{\verPresProof(VP)}$ the Verifier verifies the Issuer signature of $\TT{\SDS}$ and compares ${\HH(a_{i_j}||s_{i_j})}$ with ${L_{i_j}}$, for each $(a_{i_j},s_{i_j})\in \TT{\DA\times\DS}$. If the signature is verified and the digests ${\HH(a_{i_j}||s_{i_j})}$ match with ${L_{i_j}}$, the VP is accepted.

\subsection{Merkle Tree Mechanism}
\label{subsub:merTree}

The $\TT{\merTree}$ mechanism uses Merkle trees to create commitments $\TT{\SDS}$.

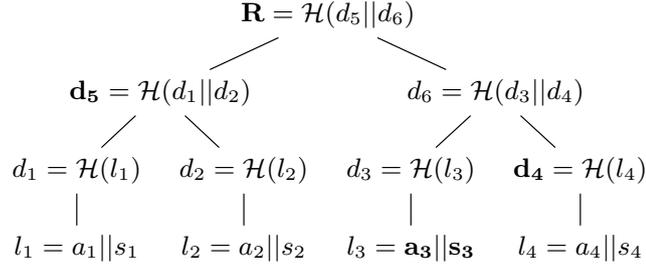
\begin{figure}
\centering
\resizebox{.58\linewidth}{!}{
\begin{forest}
 /tikz/every node/.append style={font=\footnotesize},
  [${\mathbf{R}=\HH(d_5||d_6)}$
    [${\mathbf{d_5}=\HH(d_1||d_2)}$[${d_1=\HH(l_1)}$[${l_1=a_1||s_1}$]][${d_2=\HH(l_2)}$[${l_2=a_2||s_2}$]]]
    [${d_6=\HH(d_3||d_4)}$[${d_3=\HH(l_3)}$[${l_3=\mathbf{a_3||s_3}}$]][${\mathbf{d_4}=\HH(l_4)}$[${l_4=a_4||s_4}$]]]
  ]
\end{forest}
}

\caption{Merkle tree constructed over 4 leaves. Disclosing $a_3||s_3$, their inclusion proof in $R$ is $[3,d_4,d_5]$.}
\label{fig:merkleExample}
\end{figure}

$\TT{\genIssProof(sk_{Iss},\ARI,\salt)}$: the Issuer generates one random salt $s_i$ for each attribute $a_i$, then uses their ordered concatenated pairs as leaves of a Merkle tree. The Issuer sets the $\TT{\SDS}$ equal to the Merkle tree root 
\begin{equation}
    R=\TT{getRoot}(a_1||s_1,a_2||s_2,\dots,a_m||s_m).
\end{equation}
An example of Merkle tree is given in Figure \ref{fig:merkleExample}.

To create a VP, the Holder includes the presentation payload as in column 2 of Table \ref{tab:pres}. 
The presentation proof, together with the signed commitment, also requires the Holder-generated proof, which the Holder obtains by computing the inclusion paths of the attributes that the Holder wants to disclose. 

The Verifier verifies the presentation computing $\TT{\verPresProof(VP)}$ verifying the signature of $\TT{\SDS}$ and verifying that the inclusion paths in $P$ let the Verifier reconstruct the signed root ${R}$, for each \\
$\TT{(a_{i_j},s_{i_j})\in \DA\times\DS}$.

For example, the inclusion path of the leaf $l_3$ in position 3 of the Merkle tree in Figure \ref{fig:merkleExample}, given the public root $R$, is ${[3, d_4, d_5]}$. In order to verify the inclusion of $l_3$, the Verifier computes ${d_3=\HH(l_3)}$, ${d_6=\HH(d_3||d_4)}$, and verifies that $\HH(d_5||d_6)=R$.

\section{{Selective Disclosure Signature Mechanism}}
\label{sec:sd-signatures}

\emph{Selective disclosure signatures}, following the naming in \cite{VC_imp_guide}, are a class of digital signature algorithms that enable (a) an Issuer to sign multiple attributes with a single signature, (b) a Holder to prove possession of a signature and some undisclosed attributes, generating fresh NIZKP without involving the Issuer - recall Section~\ref{subsec:NIZKP}, and (c) a Verifier to verify the validity of a disclosed subset of attributes, given only the NIZKP of knowledge of the undisclosed attributes and of an associated signature. The NIZKP created by the Holder, in the literature are also referred to as \emph{signatures of knowledge}~\cite{chase2006signatures} or \emph{signatures proof of knowledge} ~\cite{camenisch2016anonymous}.

Examples of selective disclosure signatures are $\TT{\CL}$ (Section~\ref{subsec:CL}), $\TT{BBS}$ (Section~\ref{subsec:BBS}), $\TT{\BBS}$ (Section~\ref{subsec:BBS+}) and $\TT{\PS}$ (Section~\ref{subsec:PS}), which are signature algorithms for which an ordered list of messages is input to the signature generation $\TT{\sign(sk_{Iss},(a_1,\dots,a_m))}=\sig$ and signature verification \\
$\TT{\ver(pk_{Iss},(a_1,\dots,a_m),\sig)=true/false}$.

The security of the schemes we describe below resides on the security of the selective disclosure digital signature used, as discussed in Section \ref{subsec:digsig}, and on the security of the NIZKP we will present, as discussed in Section \ref{subsec:NIZKP} and more in detail in \ref{sigmaFS}. In the next sections, before the description of each selective disclosure signature, we will mention the assumptions used to prove their security.

\paragraph{Operations in the Issuing Phase.}
The VC based on the use of selective disclosure signature algorithms as cryptographic mechanism is composed of three parts (see column 3 of Table \ref{tab:cred}):
\begin{itemize}
    \item Issuer protected header, containing the name of the cryptographic mechanism $\TT{\cm}$ i.e., the chosen selective disclosure signature scheme, and the Issuer public key $\TT{pk_{Iss}}$;
    \item Issuer payloads, containing the list of attributes $\TT{\ARI}=(a_1,\dots,a_m)$;
    \item Issuer proof, containing the \emph{selective disclosure signature} ($\TT{\SDSig}$) of the attributes in $\TT{\ARI}$, \\
    $\sigma=\TT{\sign(sk_{Iss},\ARI})$.
\end{itemize}

Therefore the function that allows the Issuer to create the Issuer proof is just $\TT{\genIssProof(sk_{Iss},\ARI)}=\TT{\sign(sk_{Iss},\ARI)}$, and the function that allows the Holder to verify it is $\TT{\verIssProof(VC)}=\TT{\ver(pk_{Iss},\ARI,\sig)}$.

\paragraph{Operations in the Presentation Phase.}
To selectively disclose some attributes of a VC to a Verifier, the Holder creates a VP (see  column 3 of Table \ref{tab:pres}) composed of:
\begin{itemize}
    \item presentation protected header, containing the name of the cryptographic mechanism $\TT{\cm}$ and the Issuer public key;
    \item presentation payload, containing the list\\
    $\TT{\DA}=(a_{i_1},\dots,a_{i_d})$ of disclosed attributes;
    \item presentation proof ${P}$, generated by the Holder executing
    $\TT{\genHolderProof(pk_{Iss},\DA,\ARI,\sig)}$, a NIZKP of the signature $\TT{\sig}$, certifying the revealed attributes in $\TT{\DA}$ and proving in zero-knowledge the knowledge of the hidden attributes in $\TT{\ARI\setminus \DA}$.
\end{itemize}
The Verifier verifies the NIZKP ${P}$ by computing the function $\TT{\verPresProof(VP)}$.

For $\TT{\CL}$, $\TT{BBS}$, $\TT{\BBS}$ and $\TT{\PS}$  we provide a 
high level description of $ \TT{\genHolderProof(pk_{Iss},\DA,\ARI,\sig)}$ and\\ $\TT{\verPresProof(VP)}$, including references to computation details omitted for brevity. 

\paragraph{$\TT{\SDS}$ vs $\TT{\SDSig}$.} 
The purpose of $\TT{\SDS}$ is to bind the attributes into an item that is subsequently signed by the Issuer. 
The Holder can perform selective disclosure by revealing $\TT{\SDS}$, the attributes to be disclosed, and a presentation proof.  
On the other hand, $\TT{\SDSig}$ simultaneously binds the attributes into an item that is itself a digital signature, certifying the authorship of the VC. To create a presentation, the Holder must not reveal $\TT{\SDSig}$, but rather derive from $\TT{\SDSig}$ a randomized proof that assures the Verifier about the 
claims. A detailed comparison between the cryptographic mechanisms that use $\TT{\SDS}$ or $\TT{\SDSig}$ is included in Section \ref{sec:secAnalys} and Section \ref{subsec:evaluation}.

\subsection{CL Signature}
\label{subsec:CL}
The CL signature scheme 
was presented by Camenish and Lysyanskaya and its security relies on the strong RSA assumption~\cite{camenisch2002signature}. 

The CL digital signature algorithm is  defined as follows \cite{IdeMix}:

\begin{description}

\item \textit{Key generation algorithm }$\TT{\keyGen}()$. 
Let ${n\leftarrow pq}$ be an $\ell_n$-bit \emph{special RSA modulus},\footnote{$n=pq$ is a special RSA modulus if $p=2p'+1$ and $q=2q'+1$ with $p',q'$ prime numbers} and choose uniformly at random quadratic residues $R_1,\dots,R_{m},S,Z$.\footnote{$q$ is a quadratic residue modulo $n$ if there exists $a\in \mathbb{Z}_n$ such that $q=a^2 \mod n$. Note that these elements depend on the public key $n$.}

Output the public key 
\begin{align}\label{eq:pk_CL}
    \TT{pk_{Iss}}=(n,R_1,\dots,R_{m},S,Z)   
\end{align}
and the secret key 
\begin{align}\label{eq:sk_CL}
    \TT{sk_{Iss}}=(p).
\end{align}

\item \textit{Signing algorithm }$\TT{genSig(sk_{Iss},\ARI)}$. 
On input the messages 
\begin{align}\label{eq:messages_CL}
    \TT{\ARI}=\{a_1,\dots,a_{m}\}, a_i\in\{0,1\}^{\ell_a},
\end{align}
and a secret key (\ref{eq:sk_CL})
 choose a random prime number ${e}\in\{0,1\}^{\ell_e}$, ${\ell_e>\ell_a+2}$, ${e>2^{\ell_e-1}}$, and a random number ${v\in \{0,1\}^{\ell_v}}$, where ${\ell_v=\ell_n+\ell_a+\ell_\emptyset}$ with ${\ell_\emptyset}$ a security parameter (e.g. ${\ell_\emptyset=80}$). 
Compute

\begin{align}
\label{eq:CL_struct}
{A\leftarrow \left(\frac{Z}{R_1^{a_1} \dots R_m^{a_m}S^v}\right)^{\frac{1}{e}}\mod n}
\end{align}
where $\frac{1}{e}$ is computed modulo $\phi(n)=(p-1)(q-1)$.
The resulting output signature is 
\begin{align}\label{eq:sig_CL}
    \sig = (A,e,v).
\end{align}

\item \textit{Verification algorithm.} $\TT{\ver (pk_{Iss},\ARI,\sig)}$. 
On input a public key (\ref{eq:pk_CL}), a set of messages (\ref{eq:messages_CL}) and a CL signature (\ref{eq:sig_CL}), check that the following holds:
\begin{align}
    Z &= A^eR_1^{a_1}\dots R_m^{a_m}S^v \mod n   \\
    a_i &\in \{0,1\}^{\ell_{a}}  \\
    e &\in [2^{\ell_e-1}+1,2^{\ell_e}-1]
\end{align}
\end{description}

These functions completely define 

$\TT{\genIssProof(sk_{Iss},\ARI)}$ which corresponds to \\
$\TT{SDSig=\sign(sk_{Iss},\ARI})=\sig=(A,e,v)\in \mathbb{Z}_n\times \{0,1\}^{\ell_e}$\\
$ \times \{0,1\}^{\ell_v}$ and $\TT{verIssProof(VC)}$.

\paragraph{VP Creation.}
At every presentation, the Holder, who possesses $(A,e,v)$ received from the Issuer, generates a new randomized signature $(A',e,v')$ from a signature by generating a random integer $r$ and computing $v'=v-re\in \mathbb{Z}$ and computing the NIZKP:
\begin{align}\label{eq:NIZKP_NIZKP}
    \begin{cases}
        &A'=AS^r \mod n\\
        &\pi\in SPK\{(e,v',\{m_i\not\in\TT{\DA}\}): \\
        &\hspace{50pt}\frac{Z}{\prod_{i\in \TT{\DA}}R_i^{a_i}}=A'^eS^{v'}\prod_{i\not\in \TT{\DA}}R_i^{a_i}\} 
    \end{cases}
\end{align}
according to the notation SPK introduced in Section~\ref{subsec:NIZKP}).
The Holder-generated proof presented above is a proof of knowledge of a signature from the Issuer and the attributes signed in it and has the following structure:

            \begin{align}
            P &= \TT{\genHolderProof(\DA,\ARI,\sig,pk_{Iss}})   \nonumber \\
            &=(A',\pi)=(A',c,\widehat{e},\widehat{v}',\widehat{a}_{i_1},\dots,\widehat{a}_{i_{(n-d)}}) \label{eq:genHolderProof_cl}
            \end{align}
with $\TT{pk_{Iss}}$ from Eq.~(\ref{eq:pk_CL}), and $\sig$ from Eq.~(\ref{eq:sig_CL}); $c\in \{0,1\}^{256}$ is the challenge of the underlying NIZKP;\footnote{Note that in this case the proof $\pi$ contains the challenge $c$ instead of the commitment $T$ as described in Figure \ref{fig:sigma_fiat_shamir}. This is an equivalent and more compact format for the NIZKP as we describe in Section \ref{subsec:tradeoffs}.} $A'\in \mathbb{Z}_n^*$ is a component of the randomized signature; 
\begin{align}
&\widehat{e}\in \{0,1\}^{\ell'_e+\ell_{\HH}+\ell_{\emptyset}+1}\\
&\widehat{v}'\in  \{0,1\}^{\ell_v+\ell_{\HH}+\ell_{\emptyset}+1}\\
&\widehat{a}_{i_1},\dots,\widehat{a}_{i_{(n-d)}} \in \{0,1\}^{\ell_a+\ell_{\HH}+\ell_{\emptyset}+1}
\end{align}
are the response values of the underlying NIZKP for linear relations.

The protocol is described in detail in Section 6.2.4. of~\cite{IdeMix}.
\paragraph{VP Verification.}

The verification algorithm\\
$\TT{\verPresProof(VP)}$ consists in $(i)$ verifying the NIZKP for linear relations to prove the Holder knows a valid undisclosed signature, and $(ii)$ verifying that the size of the received values $(\widehat{e},\widehat{a}_{i_1},\dots,\widehat{a}_{i_{(n-d)}})$ lies in the expected integer interval \cite{IdeMix} to ensure that the undisclosed attributes ${a}_{i_1},\dots,{a}_{i_{(n-d)}}$ and parameter $e$ used to build the NIZKP have the expected size.

\subsection{BBS Signature}
\label{subsec:BBS}
BBS signatures are group signatures presented in \cite{boneh2004short} and later of readapted in \cite{au2006constant,camenisch2016anonymous} to obtain a selective disclosure signature signature BBS+ that we describe in Section \ref{subsec:BBS+}. Recently Tessaro and Zhu \cite{Tessaro_Zhu_23} showed that the original BBS signature could be used to obtain a selective disclosure signature proving its security under the $q$-strong Diffie-Hellman assumption \cite{Tessaro_Zhu_23}. This signature algorithm is the object of a standardization effort from W3C and has led to an RFC draft by IRTF \cite{BBS_spec} which aims to standardize also the associated NIZKP. 

The algorithms defining the BBS signature are:
\begin{description}
    \item \textit{Set-up.} Let $\mathbb{G}_1=<g_1>,\mathbb{G}_2=<g_2>$ and $\TT\mathbb{G}_T$ be groups of prime order ${p}$, ${\mathbf{e}:\mathbb{G}_1\times \mathbb{G}_2\rightarrow \mathbb{G}_T}$ be a pairing\footnote{A pairing is a map satisfying \emph{bilinearity}, i.e. ${\mathbf{e}(g_1^x,g_2^y)=\mathbf{e}(g_1,g_2)^{xy}}$, \emph{non-degeneracy}, i.e. for each generator ${g_1\in \mathbb{G}_1, g_2\in \mathbb{G}_2}$, then ${\mathbf{e}(g_1,g_2)}$ generates ${\mathbb{G}_T}$, and \emph{efficiency} which means that the map can be efficiently computed for any input.} and ${(h_1,\dots,h_m)\in \mathbb{G}_1^{m}}$ a random vector. Set the public parameters 
    \begin{align}\label{eq:pp_BBS}
    &\pp=(p,\mathbb{G}_1,g_1,\mathbb{G}_2,g_2,G_T,\mathbf{e},h_1,\dots,h_m).
    \end{align}
    \item \textit{Key generation algorithm} $\TT{\keyGen(\pp)}$. Take a random ${x\in \mathbb{Z}_p^*}$, set 
    \begin{align}\label{eq:sk_BBS}
    \TT{sk_{Iss}}=x
    \end{align}
    and set 
    \begin{align}\label{eq:pk_BBS}
        \TT{pk_{Iss}}=w=g_2^x.
    \end{align}
    \item \textit{Signing algorithm }$\TT{genSig(sk_{Iss}}=x,\TT{\ARI)}$. On input the secret key (\ref{eq:sk_BBS}) and the messages 
    \begin{align}\label{eq:messages_BBS}
        {\TT\ARI}=(a_1,\dots,a_m)\in \mathbb{Z}_p^m,
    \end{align}  
    randomly generate $e\in \mathbb{Z}_p$ and compute 
    
    \begin{align}
    &C=(g_1\prod_{i=1}^{m}h_i^{a_i})\\
    &A=C^{\frac{1}{e+x}}.
    \label{eq:BBS_struct}
    \end{align}
    Output the pair 
    \begin{align}\label{eq:sig_BBS}
        \sigma=(A,e)\in \mathbb{G}_1\times \mathbb{Z}_p.
    \end{align}
    
    \item \textit{Verification algorithm} $\TT{\ver (pk_{Iss},\ARI,\sig)}$. On input the public key (\ref{eq:pk_BBS}), the messages (\ref{eq:messages_BBS}), and a signature (\ref{eq:sig_BBS}), set $C=g_1\prod_{i=1}^{m}h_i^{a_i}$ and check that 
    \begin{align*}
    \mathbf{e}(A,wg_2^e)=\mathbf{e}(C,g_2).
    \end{align*}
\end{description}

\paragraph{VP Creation.}
The Holder can generate a VP with\\ $\TT{\genHolderProof(\DA,\ARI,\sig,pk_{Iss})}$ whose output is obtained from the construction of a NIZKP of knowledge of the signature and the hidden attributes based on the NIZKP for linear relations. The Holder samples uniformly at random $r\in \mathbb{Z}_p$, computes:
\begin{align}
    C_D=g_1\prod_{i\in \TT{\DA}}h_i^{a_i},
\end{align}
and computes the NIZKP:
\begin{align}\label{eq:NIZKP_BBS}
    \begin{cases}
        \overline{A}=A^r    \\
        \overline{B}=C^r\overline{A}^{-e}   \\
        \pi\in SPK\{(r,e,\{a_i\not\in\TT{\DA}\}):\overline{B}=C_D^r\overline{A}^{-e}\prod_{i\not\in\TT{\DA}}h_i^{ra_i}\}
    \end{cases}
\end{align}
The function returns 
\begin{align}
P &= \TT{\genHolderProof(\DA,\ARI,\sig,pk_{Iss})}\nonumber \\
        &=(\overline{A},\overline{B},\pi)={(\overline{A},\overline{B},T,\widehat{r},\widehat{e},\widehat{a}_{i_1},\dots, \widehat{a}_{i_{m-d}})} \label{eq:genHolderProof_bbs}
\end{align}
where $\overline{A},\overline{B},T\in \mathbb{G}_1$, and all other elements lie in $\mathbb{Z}_p$. For a detailed description and the security proofs we refer to \cite{Tessaro_Zhu_23}.

\paragraph{VP Verification.}
Having received a VP from a Holder, the Verifier computes the function $\TT{\verPresProof(VP)}$, which consists in executing the verification steps of the underlying NIZKP for linear relations and verifying that the terms $\mathbf{e}(\overline{A},w)= \mathbf{e}(\overline{B},g_2)$.

\paragraph{An alternative VP construction.} 
When creating a VP from multiple VCs, the separate randomization of each attribute in each VC may be a hindrance to proving predicates such as the equality of two hidden attributes.

In \cite{BBS_spec}, the authors propose an alternative construction of the VP,\footnote{Private communication with one of the authors of~\cite{BBS_spec}.} which allows the Holder not to store the variable $C_D=g_1\prod_{i\in \TT{\DA}}h_i^{a_i}$, which is used as an element of the representation of $\overline{B}$, when she computes a proof for a VP. Instead, since $\overline{B}$ must be sent to the verifier in any case, and since $\overline{B}=C^r\overline{A}^{-e}=C_D^r \prod_{i\not\in \DA}g_i^{ra_i}\overline{A}^{-e}$ holds, then the Holder proves knowledge of a representation of $C_D$ as follows: 
\begin{align}
    C_D=\overline{B}^{r^{-1}}\prod_{i\not\in \TT{\DA}}h_i^{-a_i} \overline{A}^{er^{-1}}.
\end{align}
Therefore the Holder computes:
\[\pi\in SPK\{(r,e,\{a_i\not\in\TT{\DA}\}):C_D=\overline{B}^{r^{-1}}\prod_{i\not\in \TT{\DA}}h_i^{-a_i} \overline{A}^{er^{-1}}\}.\]

This alternative algorithm and another variant is described in the appendix of a recent update\footnote{\url{https://eprint.iacr.org/2023/275} updated on 2023-12-09} of the paper presented at Eurocrypt 2023 by Tessaro and Zhu \cite{Tessaro_Zhu_23}.
\subsection{BBS+ Signature}
\label{subsec:BBS+}

The BBS+ signature was presented by Au et al.~\cite{au2006constant} as a provably secure extension to BBS group signatures~\cite{boneh2004short} and improved by Camenisch et al.~\cite{camenisch2016anonymous}. Its security relies on the $q$-strong Diffie-Hellman assumption~\cite{camenisch2016anonymous}.
The digital signature BBS+ is defined by the following algorithms:

\begin{description}
    \item \textit{Set-up.} Let $\mathbb{G}_1=<g_1>,\mathbb{G}_2=<g_2>$ and $\TT\mathbb{G}_T$ be groups of prime order ${p}$, ${\mathbf{e}:\mathbb{G}_1\times \mathbb{G}_2\rightarrow \mathbb{G}_T}$ be a pairing and ${(h_0,\dots,h_m)\in \mathbb{G}_1^{m+1}}$ a random vector. Set the public parameters 
    \begin{align}
    \label{eq:pp_BBS+}
    &\pp=(p,\mathbb{G}_1,g_1,\mathbb{G}_2,g_2,G_T,\mathbf{e},h_0,\dots,h_m).
    \end{align}
    \item \textit{Key generation algorithm }$\TT{\keyGen(\pp)}$. Sample uniformly at random a random ${x\in \mathbb{Z}_p^*}$, set 
    \begin{align}\label{eq:sk_BBS+}
    \TT{sk_{Iss}}=x,
    \end{align}
    then set 
    \begin{align}\label{eq:pk_BBS+}
        \TT{pk_{Iss}}=w=g_2^x. 
    \end{align}
    \item \textit{Signing algorithm }$\TT{genSig(sk_{Iss}}=x,\TT{\ARI)}$. On input the secret key (\ref{eq:sk_BBS+}) and the messages 
    \begin{align}\label{eq:messages_BBS+}
        {{\TT\ARI}=(a_1,\dots,a_m)\in \mathbb{Z}_p^m},
    \end{align}  
    randomly generate $e,s\in \mathbb{Z}_p$, compute 
    \begin{align}
        \label{eq:BBS+_struct}
        &C=g_1h_0^s\prod_{i=1}^{m}h_i^{a_i}\\
        & A=C^{\frac{1}{e+x}}.
    \end{align}
    Output the triple 
    \begin{align}\label{eq:sig_BBS+}
    \sigma=(A,e,s).
    \end{align}
    \item \textit{Verification algorithm} $\TT{\ver (pk_{Iss},\ARI,\sig)}$. On input the public key (\ref{eq:pk_BBS+})
    messages (\ref{eq:messages_BBS+}), and a signature (\ref{eq:sig_BBS+}), check that the following holds: 
    \begin{align}\label{eq:BBS+_ver_issuer_proof}
        \mathbf{e}(A,wg_2^e)=\mathbf{e}(g_1h_0^s\prod_{i=1}^{m}h_i^{a_i},g_2).
    \end{align}
\end{description}

These algorithms define the functions
$\TT{\genIssProof(sk_{Iss},\ARI})$ which corresponds to\\
$\TT{SDSig=\sign(sk_{Iss},\ARI})= \sig=(A,e,s)\in \mathbb{G}_1
\times \mathbb{Z}_p^2 $, and $\TT{\verIssProof(VC)}$.

We have presented the algorithms as described by Au et Al. \cite{au2006constant}. In the paper from Camenish et Al. \cite{camenisch2016anonymous} the authors include also the elements $h_0,\dots,h_m$ as part of the public key, which they write as $(w=g_2^x,h_0,\dots,h_m)\in \mathbb{G}_2\times\mathbb{G}_1^{m}$.

\paragraph{VP Creation.}
The Holder can generate a VP proof with $\TT{\genHolderProof(\DA,\ARI,\sig,pk_{Iss})}$, whose output is obtained from the construction of a NIZKP of knowledge of the signature and the hidden attributes based on the NIZKP for linear relations. First, the Holder randomly generates ${r_1\in\mathbb{Z}_p^*}$ and ${r_2\in \mathbb{Z}_p}$. The Holder then sets

\begin{align}
    & {r_3=\frac{1}{r_1}\mod p}\\ 
    & {s'=s-r_2 r_3 \mod p}
\end{align}
and computes the NIZKP:
\begin{align}\label{eq:NIZKP_BBS+}
    \begin{cases}
        &{A'=A^{r_1}}\\
        &{\overline{A}=A'^{-e}C^{r_1}(=A'^x)} \\
        & {d=C^{r_1}h_0^{-r_2}}\\
        &\pi\in SPK\{(e,r_2,r_3,s',\{a_i\not\in\TT{\DA}\}):\\
        &\hspace{50pt}\frac{\overline{A}}{d}=A'^{-e}h_0^{r_2}\wedge\\ 
        &\hspace{50pt}g_1\prod_{i\in\TT{\DA}}h_i^{a_i}=d^{r_3}h_0^{-s'}\prod_{i\not\in\TT{\DA}}h_i^{-a_i}\}
    \end{cases}
\end{align}
The proof is then computed as
\begin{align}
P &= \TT{\genHolderProof(\DA,\ARI,\sig,pk_{Iss})}\nonumber \\
        &=(A',\overline{A},d,\pi)\\
        &=(A',\overline{A},d,T_1,T_2,\widehat{e},\widehat{r}_2,\widehat{r_3}, \widehat{s}',\widehat{a}_{i_1},\dots, \widehat{a}_{i_{m-d}}) \label{eq:genHolderProof_bbs+}
\end{align}
where $A',\overline{A},d,T_1,T_2\in \mathbb{G}_1$, and all other elements lie in $\mathbb{Z}_p$. For a detailed description we refer to \cite{camenisch2016anonymous}.

\paragraph{VP Verification.}
Having received a VP from a Holder, the Verifier computes the function $\TT{\verPresProof(VP)}$, which consists in executing the verification steps of the underlying NIZKP for linear relations and verifying that the terms $A'\ne 1_{\mathbb{G}_1}$ computing $\mathbf{e}(A',w)=\mathbf{e}(\overline{A},g_2)$.

\paragraph{Adapting the NIZKP for VC based on BBS to NIZKP for VC based on BBS+.}
In order to emphasize the differences between BBS and BBS+ signatures on messages $a_1,\dots,a_m$, we describe the BBS+ signature starting from the BBS signature.
\begin{itemize}
    \item the public parameters $\pp$ of BBS+ (Eq. \ref{eq:pp_BBS+}) are the same as the ones of BBS (Eq. \ref{eq:pp_BBS}) with an extra random element $h_0\in\mathbb{G}_1$;
    \item the variable $C$ computed to generate a BBS signatures is $C=g_1\prod_{i=1}^mh_i^{a_i}$ (Eq. (\ref{eq:BBS_struct})), while for BBS+ signatures (we rename $C$ as $C'$ to distinguish it from the one used in BBS) the signer must generate at random $s\in\mathbb{Z}_p$ and compute $C'=g_1h_0^s\prod_{i=1}^mh_i^{a_i}=C h_0^s$ (Eq. (\ref{eq:BBS+_struct}));
    \item the BBS signature is given by $(A,e)=(C^{\frac{1}{x+e}},e)$ (Eq. (\ref{eq:sig_BBS})) while the BBS+ signature is given by $(A,e,s)=(C'^{\frac{1}{x+e}},e,s)$.
\end{itemize}

Once highlighted these differences between the two, it is clear that a BBS+ signature $(A,e,s)$ over messages $(a_1,\dots,a_m)$ w.r.t. the public parameters $$\pp=(p,\mathbb{G}_1,g_1,\mathbb{G}_2,g_2,G_T,\mathbf{e},h_0,\dots,h_m)$$ can be univocally turned into a BBS signature $(A,e)$ over the messages $(s,a_1,\dots,a_m)$ w.r.t. exactly the same public parameters $\pp$.
Therefore, proving knowledge of a BBS+ signature $(A,e,s)$ and of some hidden attributes $H=\{a_i\not\in\DA\}$, without revealing it, would be equivalent to prove knowledge of the univocally determined BBS signature $(A,e)$ and of the same attributes $H$ to which we will add $s$.

This means that the NIZKP used for BBS signatures (Eq. (\ref{eq:NIZKP_BBS})) can be used also to prove knowledge of a BBS+ signature. The idea is the following: turn the BBS+ signature into the uniquely determined BBS signature as described above, then prove knowledge of the derived BBS signature and of the hidden attributes considering that hiding $s$ is mandatory.

\subsection{PS Signature}
\label{subsec:PS}
The Pointcheval-Sanders (PS) signature is secure under the LRSW assumption~\cite{pointcheval2016short}.The PS signature is defined by the following algorithms.

\begin{description}
    \item \textit{Set-up}
    Let $\mathbb{G}_1=<g_1>$, $\mathbb{G}_2=<g_2>$, and $\mathbb{G}_T$ be groups of prime order ${p}$, and ${\mathbf{e}:\mathbb{G}_1\times \mathbb{G}_2\rightarrow \mathbb{G}_T}$ be a pairing. Set the public parameters
    
    \begin{align}
    \pp=(p,\mathbb{G}_1,g_1,\mathbb{G}_2,g_2,G_T,\mathbf{e}).
    \end{align}
    
    \item \textit{Key generation algorithm }$\TT{\keyGen(\pp)}$. Take a random vector
    \begin{align}\label{eq:sk_PS}
        \TT{sk_{Iss}}=(x,y_1,\dots,y_m)\in \mathbb{Z}_p^{m+1},
    \end{align}
    then set
    \begin{align}\label{eq:pk_PS}
        \TT{pk_{Iss}}&=(X,Y_1,\dots,Y_m)\\&= (g_1^x,g_1^{y_1},\dots,g_1^{y_m})\in \mathbb{G}_1^{m+1},
    \end{align}
    \item \textit{Signing algorithm }$\TT{genSig(sk_{Iss}},\TT{\ARI)}$. On input the secret key (\ref{eq:sk_PS}) 
    and the messages 
    \begin{align}\label{eq:messages_PS}
        {{\TT\ARI}=(a_1,\dots,a_m)\in \mathbb{Z}_p^m},
    \end{align}
    randomly generate $h\in \mathbb{G}_2^*$ and compute
    \begin{align}\label{eq:sig_PS}
        \sig=(\sig_1,\sig_2)=(h,h^{x+\sum_{j=1}^m{y_ja_j}})\in \mathbb{G}_2^2
    \end{align}
    \item \textit{Verification algorithm} $\TT{\ver (pk_{Iss},\ARI,\sig)}$. On input a public key (\ref{eq:pk_PS}),
    messages (\ref{eq:messages_PS})
    , and a signature 
    (\ref{eq:sig_PS}), check that both (\ref{eq:ver_sig_PS_1}) and (\ref{eq:ver_sig_PS_2}) hold:
    \begin{align}\label{eq:ver_sig_PS_1}
        h&\ne 1_{\mathbb{G}_1}   \\\label{eq:ver_sig_PS_2}
        \mathbf{e}(g_1,\sigma_2)&=\mathbf{e}(X\prod_{i=1}^m{Y_i^{a_i}},\sig_1).
    \end{align}
    
\end{description}

\paragraph{VP Creation.}
The Holder can generate a VP with $\TT{\genHolderProof(\DA,\ARI,\sig,pk_{Iss})}$ whose output is obtained from the construction of a NIZKP of knowledge of the signature and the hidden attributes based on the NIZKP for linear relations applied to a randomized signature.
In particular, the Holder computes a signature of the messages $(a_1,\dots,a_m,t)$, where $t\in \mathbb{Z}_p$ is a random message associated to the dummy public key $Y_{m+1}=g_1$, i.e. $(h,h^{x+(\sum_{j=1}^my_ja_j)+t})\in \mathbb{G}_2^2$, then randomizes it by picking a random $r\in \mathbb{Z}_p$  and computing:

    \begin{align}
    \begin{cases}
    &\sigma'=(\sigma'_1,\sigma'_2)=(h^r,h^{r(x+(\sum_{j=1}^my_ja_j)+t)})\\
        &\pi\in SPK\biggl\{(t,\{m_i\not\in \TT{\DA}\}):
        \mathbf{e}(g_1,\sigma'_1)^t\prod_{i\not\in\TT{\DA}}\mathbf{e}(Y_i,\sigma'_1)^{m_i}\\
        &\hspace{10pt}=\mathbf{e}(g_1,\sigma'_2)\left(\mathbf{e}(X,\sigma'_1)\prod_{i\in \TT{DA}}\mathbf{e}(Y_i,\sigma'_1)^{m_i}\right)^{-1}\biggr\}.
        \end{cases}
    \end{align}

and returns to the Verifier the tuple
\begin{align}
P &= \TT{\genHolderProof(\DA,\ARI,\sig,pk_{Iss})}\nonumber \\
        &=(\sig_1',\sig_2',\pi)=(\sig_1',\sig_2',T,\widehat{t},\widehat{a}_{i_1},\dots,\widehat{a}_{i_{m-d}})
\end{align}
where $\sig_1',\sig_2'\in \mathbb{G}_2$, $T\in \mathbb{G}_T$ and the other elements are in $\mathbb{Z}_p$
\cite{Pointcheval_Sanders_18}.

\paragraph{VP Verification}
Having received a VP from a Holder, the Verifier computes the function $\TT{\verPresProof(VP)}$, which consists in executing the verification steps of the underlying NIZKP for linear relation.

As we will show in the next paragraph it is possible to avoid to perform computations in $\mathbb{G}_T$ and avoid to compute so many pairings as one would expect by looking at the SPK described above.

\paragraph{A more practical VP construction.} The algorithm presented above for the creation of the Holder-generated proof requires the Holder to perform computations in $\mathbb{G}_T$, the codomain of the pairing $\mathbf{e}:\mathbb{G}_1\times\mathbb{G}_2\to \mathbb{G}_T$. However, this can be avoided according to the implementation proposed in Ursa.\footnote{\url{https://docs.rs/ursa/}} In fact, it is possible to observe that, by the bilinearity of $\mathbf{e}$, it holds that 
$$\mathbf{e}(g_1,\sigma'_1)^t\mathbf{e}(X,\sigma'_1)\prod\mathbf{e}(Y_i,\sigma'_1)^{a_i}=\mathbf{e}(g_1^tX\prod Y_i^{a_i},\sigma'_1),$$
so the Holder can send to the Verifier 
\begin{align}\label{eq:NIZKP_PS}
    \begin{cases}
        &\sigma'=(\sigma'_1,\sigma'_2)=(h^r,h^{r(x+(\sum_{j=1}^my_ja_j)+t)})\\
        &J=g_1^t\prod_{i\not\in\TT{\DA}}Y_i^{a_i}\\
        &\pi=SPK\left\{\left(t,\{a_i\not\in\TT{DA}\}\right): J=g_1^t\prod_{i\not\in\TT{DA}}Y_i^{a_i}\right\}=\\
        &\hspace{10pt}=(T,\widehat{t},\widehat{a}_{i_1},\dots,\widehat{a}_{i_{m-d}})\in\mathbb{G}_1\times \mathbb{Z}_p^{m-d+1}
    \end{cases}
\end{align}
where $r,t\in\mathbb{Z}_p$ are the random elements used to randomize the signature, $d$ is the number of disclosed attributes and $m$ the number of attributes in the VC. In this way, the Holder-generated proof has the following form: \begin{align}
P &= \TT{\genHolderProof(\DA,\ARI,\sig,pk_{Iss})}\nonumber \\
        &=(\sig_1',\sig_2',J,\pi)\\
        &=(\sigma_1',\sigma_2',J,T,\widehat{t},\widehat{a}_{i_1},\dots,\widehat{a}_{i_{m-d}})\in \mathbb{G}_2^2\times\mathbb{G}_1^2\times \mathbb{Z}_p^{m-d+1}
        \label{eq:genHolderProof_ps}
\end{align}
To verify $\pi$, the Verifier can compute $J'=JX\prod_{i\in\TT{DA}}Y_i^{a_i}$ and check that $\mathbf{e}(\sigma'_1,J')\mathbf{e}({\sigma'}_2^{-1},g_1)=1_{\mathbb{G}_{T}}$.

In this way all the computations are performed in $\mathbb{G}_1$ which is the group in which computations are more efficient among the ones involved in the pairing definition. Also the Holder does not have to compute any pairing and the number of pairing computations performed by the Verifier is reduced to two.

\subsection{Efficiency and Trust on Issuer Set-up Domain Parameters}
\label{subsec:set-up}



For BBS and BBS+ signatures, the public parameters output by \texttt{\setUp($\lambda$)} that give structure to a VC and are used to generate and verify VCs and VPs may be generated from a seed in a manner that is not confidential. It is possible to reduce the size of data required for verification by requiring the Verifier to reconstruct the public parameters from the seed. This is not true for CL and PS digital signatures, where parameters must be generated by the Issuer in the key generation algorithm and are part of the Issuer's public key.

In addition, BBS and BBS+ public parameters may be provided by a trusted third party. This may enable the re-use of the same set of public parameters by multiple Issuers, e.g., for the same kind of VC.

If the dimension of the public key is not a concern or it is preferable to require each Issuer to perform its own setup, the public parameters of Eq. (\ref{eq:pp_BBS}) and Eq. (\ref{eq:pp_BBS+}) can be included in the public key; indeed, this is how they appear in \cite{camenisch2016anonymous,hesse2023bind,anoncreds_2_spec_Ursa}.
\begin{itemize}
    \item CL: the quadratic residues $R_1,\dots,R_m,S,Z\in \mathbb{Z}_n$ are needed to generate the $A$ component of the signature according to Eq. (\ref{eq:CL_struct}). These elements are qua-\\
    dratic residues modulo $n=pq$, therefore must be computed according to the secret key (Eq. (\ref{eq:sk_CL})) and must necessarily be part of the Issuer public key (Eq. (\ref{eq:pk_CL})).
    
    \item BBS (BBS+): $h_1,\dots,h_m $ ($h_0,\dots,h_m$) are needed to generate the $A$ component of the signature according to Eq. (\ref{eq:BBS_struct}) (Eq. (\ref{eq:BBS+_struct})).
    These are random elements of the group $\mathbb{G}_1$ and do not depend on the Issuer secret key in Eq. (\ref{eq:sk_BBS}) (Eq. (\ref{eq:sk_BBS+})). As proposed in \cite{BBS_spec,Tessaro_Zhu_23}, these parameters may be generated either by the Issuer or by a trusted third party by using a hash-to-curve function that maps a seed to a set of random elements in $\mathbb{G}_1$ of the required cardinality. For security reasons it must be infeasible to compute the discrete logarithm of $h_i$, therefore these random elements can not be generated by picking random scalars $z_1,\dots,z_m$ and computing $h_i=g_1^{z_i}$. 
    \item PS: $(X,Y_1,\dots,Y_m)$ are needed to give structure to the element of $\mathbb{G}_1$ in the right-hand side of Eq. (\ref{eq:ver_sig_PS_2}), whose validity is part of the signature verification process. In contrast with the approach proposed by BBS and BBS+, the signer must know the discrete logarithm $x,y_1,\dots,y_m$ of $X,Y_1,\dots,Y_m$ w.r.t. $g_1$ to compute their analogue w.r.t. any basis $h\in \mathbb{G}_2$ during the signing process. In fact, the PS signature (Eq. (\ref{eq:sig_PS})) can be rewritten as $(h,h^x\prod_{i=1}^m (h^{y_i})^{a_i})$. For this reason $(X,Y_1,\dots,Y_m)$ are part of the PS public key (Eq. (\ref{eq:pk_PS})) and their discrete logarithms w.r.t. $g_1$ are the secret key (Eq.(\ref{eq:sk_PS})).
\end{itemize}


\section{Solution Design Analysis}
\label{sec:secAnalys}
To assess the maturity of options, we consider their standardization (Section~\ref{subsec:standardization}), cryptographic agility (Section~\ref{subsec:cryptographic_agility}) and quantum safety (Section~\ref{subsec:quantum}).

\subsection{Standard Maturity}
Standardization is important for cryptographic protocols to ensure expert vetting of correctness, security, and other properties claimed, as well as to promote interoperability as encouraged e.g., by the proposed Interoperable Europe Act~\cite{interoperable_europe_act_proposal}. Cryptographic agility~\cite{cryptographic_agility} ``is achieved when a protocol can easily migrate from one algorithm suite to another more desirable one, over time''~\cite{RFC_7696}. The need to transition between cryptographic algorithms and key lengths has been steadily gaining importance, e.g., replacing older versions of the Secure Hash Algorithm, and preparing for quantum computing~\cite{NIST_SP_800-131A}.

We observe how each mechanism supports privacy and offline features with regards to presentation unlinkability (Section~\ref{subsec:unlinkability}), and briefly discuss the advantages of predicate proofs (Section~\ref{subsec:predicate_proofs}). We compare the computation speed of each function described in Section~\ref{sec:selectiveDisc} 
, and the size of presentation elements of each mechanism (Section~\ref{subsec:evaluation}); we also note some trade-offs made by implementations to balance performance between these measures (Section~\ref{subsec:tradeoffs}). Finally, we describe how it is possible to perform threshold issuance of the VCs based on the cryptographic mechanisms we have described (Section~\ref{subsec:cred_aggr}). Our assessment is summarized in Section~\ref{subsec:assessment_summary}.

\subsubsection{Standardization}
\label{subsec:standardization}

\texttt{\comList} is the only mechanism featured in official standards: it is enabled by design in ISO 18013-5
~\cite{ISO_18013-5}, and it is the basis for the IETF draft SD-JWT~\cite{SD-JWT}. Both are considered mandatory for the European digital identity wallet~\cite{EUDI_ARF} developed in the context of the revised eIDAS regulation~\cite{european2021regulation}.

\texttt{\merTree} has been proposed in~\cite{jwp4mertree} as a possible mechanism for JSON Web Proof (JWP)~\cite{JWP} - a proposed container format for VCs and VPs that aims to be agnostic to the proof mechanism, currently an IETF draft on the Standards Track. \texttt{\merTree} also appears in the experimental Certificate Transparency 2.0 proposal~\cite{RFC_9162}.

The \texttt{\bbs} (previously \texttt{\bbs+}) specification~\cite{BBS_spec} is an IRTF draft. \texttt{\PS} and \texttt{\CL} signatures are not specified independently, but \texttt{\CL} appear as part of the Identity Mixer~\cite{IdeMix} and Hyperledger Ursa~\cite{anoncreds_1_spec_Ursa} anonymous credentials protocols.

ETSI Technical Report 119 476~\cite{ETSI_SD} gives recommendations on issuing, storage, and presentation of attestations under eIDAS2 in the 
form of ISO m\textsc{dl} and/or SD-JWT, with a view towards selective disclosure and unlinkability.

\subsubsection{Cryptographic Agility}
\label{subsec:cryptographic_agility}

\texttt{\comList} and \texttt{\merTree} offer the greatest agility: any cryptographic hash function can be used to construct them, and any digital signature can be chosen to sign the hash list or tree root.

\texttt{\bbs}, \texttt{\BBS}, and \texttt{\PS} signatures can in theory be based on any pairing-friendly curve, of which several have been identified~\cite{pairing-friendly_curves} up to 256-bit security, and any correspondingly secure cryptographic hash function. Cipher suites have been drafted~\cite{BBS_spec}.

The idemix specification~\cite{IdeMix} for anonymous credentials with \texttt{\CL} signatures contains a default value for 14 parameters and 7 ``constraints which parameter choices must satisfy to ensure security and soundness'' (Tables 2 and 3 therein), and it is left to the reader to adjust these as required. The default RSA modulus is 2048 bits, which corresponds to only 112 bits of security. Other than increasing the prime factor length, it is non-trivial to establish how parameters should change to increase the security level of the scheme as a whole. 

The Ursa library also defaults to 2048-bit modulus; the Ursa specification~\cite{anoncreds_1_spec_Ursa} lists individual parameter values scattered throughout, including 1536-bit RSA factors, but it is left to the reader to gather the information, to modify the source code, and to assume that all other parameters have been set to meet the same level of security.

\subsubsection{Quantum Safety}
\label{subsec:quantum}
As noted in~\cite{ETSI_SD}, cryptographic mechanisms based on hiding commitment can be instantiated using one of the post-quantum digital signature algorithms selected for standardization by NIST: CRYSTALS-Dilithium \\
\cite{lyubashevsky2020crystals}, FALCON \cite{fouque2018falcon}, or \sphincs \cite{bernstein2019sphincs+}. The use of post-quantum signatures makes these cryptographic mechanisms quantum resistant as well, as long as the hiding commitment scheme also satisfies this property - 
in particular, for those in Section \ref{sec:hiding-commitment}, as long as the cryptographic hash function 
used to create the list of commitments or the Merkle tree remains secure.

It should be noted that the three above algorithms were selected for standardization in 2022, but at the time of writing the standardization process is still in the draft stage, and some changes have been proposed in the current draft FIPS 204 and 205~\cite{FIPS_204,FIPS_205}.

For what concerns the quantum resistance of selective disclosure signatures, the algorithms described in Section \ref{sec:sd-signatures} rely on assumptions that do not hold in a post-quantum setting. Lattice-based cryptography schemes have been proposed very recently~\cite{jeudy2022lattice,bootle2023framework,Blazy_Chevalier_Renaut_Ricosset_Sageloli_Senet_23}, but there are no complete libraries available yet to make a full comparison with other schemes described here.

\subsection{Supported Features}
We evaluate for each cryptographic mechanism how they support features that are relevant to the design of practical privacy preserving VCs, namely the ability to create unlinkable VPs (Section \ref{subsec:unlinkability}), the ability to include predicate proofs in the VPs (Section \ref{subsec:predicate_proofs}) and the support for threshold credential Issuance that allow multiple Issuers to issue a single VC to a Holder (Section \ref{subsec:cred_aggr}).

\subsubsection{Presentation Unlinkability}
\label{subsec:unlinkability}

Unlinkability can be defined as ensuring that ``no correlatable data are used in a digitally-signed payload''~\cite{VC_data_integrity}. Sources of correlation include the signature itself and long-term identifiers, such as the credential subject, a credential identifier, revocation status information etc. Guaranteeing this property goes beyond selective disclosure only; here we focus on signature-based correlation.

\paragraph{Hiding Commitment.}

Since the Presentation Proof of a VP contains the 
issuer-signed commitment included in the associated VC, this identifier links each VP uniquely to one VC, and therefore to its Holder. This means that the Holder should use always different VCs to generate new VPs, therefore in the issuing phase the Issuer must provide the Holder with several distinct versions of the same VC where a distinct version of VC is built including the same set of attributes hidden using different salts. 

When all the distinct versions of the same VC have been used, the Issuer must produce and send new ones to the Holder. There must therefore be an available channel between the Issuer and the
Holder device storing the VCs that guarantees ready access to brand new VCs that can be used to create unlinkable presentations.

\paragraph{Selective Disclosure Signatures.} 
As summarized in Table \ref{tab:pres}, the presentation proof contains only the Holder-generated proof, which is a randomized element.

Given a VC, the Holder can create a new presentation proof each time that is indistinguishable from random, and therefore cannot be correlated to other VPs. The Holder can use the same VC multiple times; therefore, interaction with the Issuer is required only when requesting a new credential or renewing an expired one.

\subsubsection{Predicate Proofs}
\label{subsec:predicate_proofs}
In some use cases, there may be an interest in asking a question (``predicate'') about an attribute, without disclosing the attribute itself. For instance, a Verifier may need to know whether an m\textsc{dl} subject's age is over some threshold \texttt{NN}, or in some range, without needing to know their full date of birth. This feature would enhance privacy and follow the data minimization principle. We discuss how it is possible to create predicate proofs for hiding commitment based cryptographic mechanisms and for selective disclosure signature cryptographic mechanisms.

\paragraph{Hiding commitments.}
The \texttt{\comList} mechanism used in m\textsc{dl} allows the Issuer to create range proofs only by treating them as individual attributes; for instance, in the AAMVA m\textsc{dl} implementation guidelines~\cite{AAMVA_mDL_implementation_guidelines} Issuers must identify every likely threshold value in their jurisdiction and encode a separate attribute \texttt{age\_} \texttt{over\_NN=True} or \texttt{False} for each \texttt{NN}.

The disadvantages of implementing this feature with this mechanism are: (a) an increased size of every VC and VP, (b) requiring the Issuer to keep track of when each threshold is crossed to issue a new VC, (c) interoperability issues - a Verifier may not find all the same thresholds represented every separate jurisdiction for the same VC type (e.g., age above 18, 21, 65, etc), (d) mistakes are easy to make and hard to spot, e.g., in a long list of individual and unrelated entries it would be possible to enter e.g., \texttt{age\_over\_18=False} and \texttt{age\_over} \texttt{\_21=True}. All hiding commitment based \texttt{\cm} including \texttt{\merTree} suffer the same disadvantages.

It is possible to create range proofs using hash functions using a protocol called HashWire \cite{chalkias2021hashwires}. HashWire is an optimization of the technique introduced by Rivest and Shamir in PayWord \cite{rivest1996payword}. To create a commitment to an integer $k$, the Issuer generates a random string $r$ and computes the commitment $c=\HH^k(r)$, namely $k$ repeated iterations of the hash function $\HH$. The Issuer reveals the random string $r$ and the integer $k$ to the Holder, who can prove to a Verifier that $k\ge t$, for a given threshold $t$, by sending the proof $\pi=\HH^{k-t}(r)$. The Verifier considers the proof $\pi$ valid if $\HH^t(\pi)=c$. This method is both more compact and less error-prone than a list of unrelated statements.

\paragraph{Selective disclosure signatures} By contrast, selective disclosure signatures enable the Holder to build NIZKP of 
predicates about the attributes included in the VC without prior involvement of the Issuer. For example, \emph{range proofs} and \emph{set membership proof}s \cite{camenisch2008rangeset} allow the Prover to prove that an attribute ${a}$ lies within a range ${v<a<u}$, or in a given set of values $\mathcal{V}$, i.e. $a\in \mathcal{V}$, respectively. Examples of predicate proofs for the \texttt{\CL} mechanism can be found in Section 6 of~\cite{IdeMix}.

\subsubsection{Support for Threshold Credential Issuance}
\label{subsec:cred_aggr}
VC ecosystems are initially designed for individual Issuers issuing VCs to Holders; however, this requires all trust to be placed on individual Issuers, which constitute a single point of failure. This problem can be mitigated by having the secret key shared among multiple Issuers, and by designing threshold digital signatures so that they may agree in order to create VCs.

An $(n,t)$-threshold signature scheme allows a group of $n$ signers to create a digital signature only if $t$ members of the group agree to sign the message.

Threshold signature schemes are often designed generalising
standard digital signature schemes. In fact, a desirable property of some threshold signature schemes is that the resulting signature has exactly the same structure of the signature they generalise, so that the verification algorithm remains unchanged.

For the threshold issuance of hiding commitment based credentials, any threshold digital signature scheme may be used since the signed commitment is always revealed. Examples of threshold signature schemes are the threshold version of EdDSA~\cite{battagliola2023provably}, ECDSA~\cite{gennaro2016threshold}, and Schnorr signature~\cite{crites2023fully}.

Threshold versions have been proposed for the \PS~signature~\cite{camenisch2020short} and for the \BBS~signature~\cite{doerner2023threshold}, which also applies to the \bbs~signature with some simple modifications. Since the signatures produced by the Issuers have the same structure as the one they generalise, the protocols for the creation and verification of VP also remain unchanged. To the best of our knowledge, no threshold version of CL has been proposed.

\section{Experimental Evaluation}
\label{subsec:evaluation}

Our use case of interest is a proximity flow for EUDI Wallets, in which the Holder and Verifier are physically close and the attestation exchange and disclosure occurs using proximity protocols - e.g., NFC, Bluetooth, QR-Codes. The Holder, Verifier, or both may also be offline. A concrete example involves checking a mobile driving license (m\textsc{dl}), as considered in Section~\ref{intro}. 
In this scenario, the Holder device may be resource-constrained in both computational capability and presentation exchange bandwidth; we therefore measure the speed of computation, particularly \texttt{\genHolderProof}, in Section~\ref{subsec:speed} and the size of VP elements for each \texttt{\cm} in Section~\ref{subsec:size}.
Based on ISO/IEC 18013-5~\cite{ISO_18013-5}, in which an m\textsc{dl} consists of $11$ mandatory and $22$ optional attributes, we use credentials with $n_a \leq33$ total attributes.

\subsection{Experimental Set-Up}
\label{subsec:evaluation_set-up}

\paragraph{Security level.}
To ensure a fair comparison, we aim for an equivalent level of security of 128 bits in all tested mechanisms - see SP 800-57~\cite{NIST_SP_800-57r5}, Table 2. Enforcing this common security level is non-trivial, as mentioned in Section \ref{subsec:cryptographic_agility}. This level is achieved by \bbs, \BBS, and \PS~over BLS12-381, \CL~with 3072-bit RSA modulus, EdDSA over ed25519, and the post-quantum signature parameter sets Falcon-512, Dilithium2, and \sphincs-SHA2-128f. Several \sphincs~sets meet the same security level, but the chosen one is optimized for speed, to the detriment of size.

\paragraph{Libraries.}

We use the Hyperledger Ursa\footnote{\url{https://docs.rs/ursa/}} library for \PS~and \CL~signature performance. As noted in Section \ref{subsec:cryptographic_agility}, we had to modify the \CL~implementation to achieve the required security level.

For \bbs~and \BBS, we test docknetwork\footnote{\url{https://github.com/docknetwork/crypto}} since they implement both versions, so differences in measurements can be attributed with greater confidence to the algorithm rather than implementation differences. They are also the recently most used BBS crate\footnote{\url{https://crates.io/search?q=bbs\%20signature\&sort=recent-downloads}} after Ursa at the time of writing, and an active project.

For all proofs in \texttt{merTree} mechanisms we use SHA-256 and rs\_merkle\footnote{\url{https://docs.rs/rs_merkle/}}. For \texttt{merTree} digital signatures we test EdDSA over ed25519 using the popular rust crate ed25519-dalek\footnote{\url{https://docs.rs/ed25519-dalek/}}, and three PQ signature standardization candidates implemented in Open Quantum Safe (OQS).\footnote{\url{https://github.com/open-quantum-safe/liboqs}}

\paragraph{Processors.}
In order to test performance on both desktop PCs and constrained devices with ARM CPUs more closely resembling mobile phones - our use case for Holder devices - experiments are run on AMD Ryzen 7 5800X, raspberry pi 3B+ 1GB RAM, and pi 4B 4GB RAM. 
\subsection{Speed}
\label{subsec:speed}
\begin{figure}[!ht]

      \centering
      \includegraphics[width=.7\textwidth]{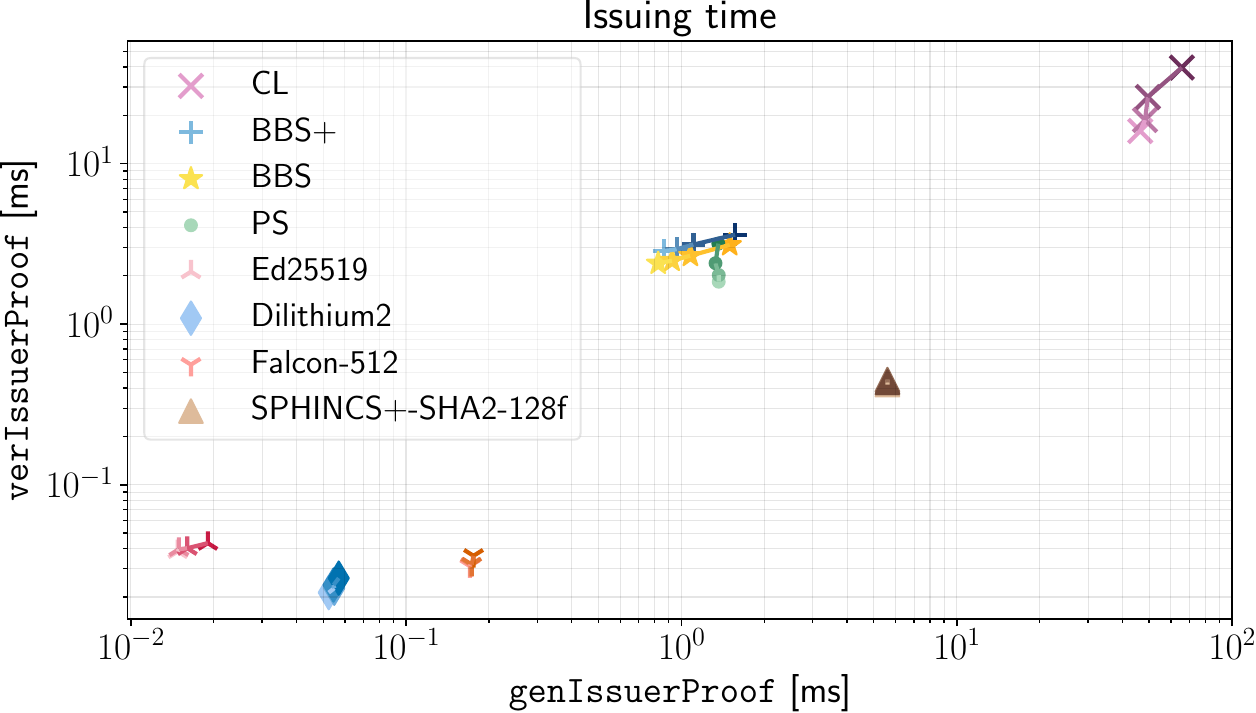}
      \centering
      \includegraphics[width=.7\textwidth]{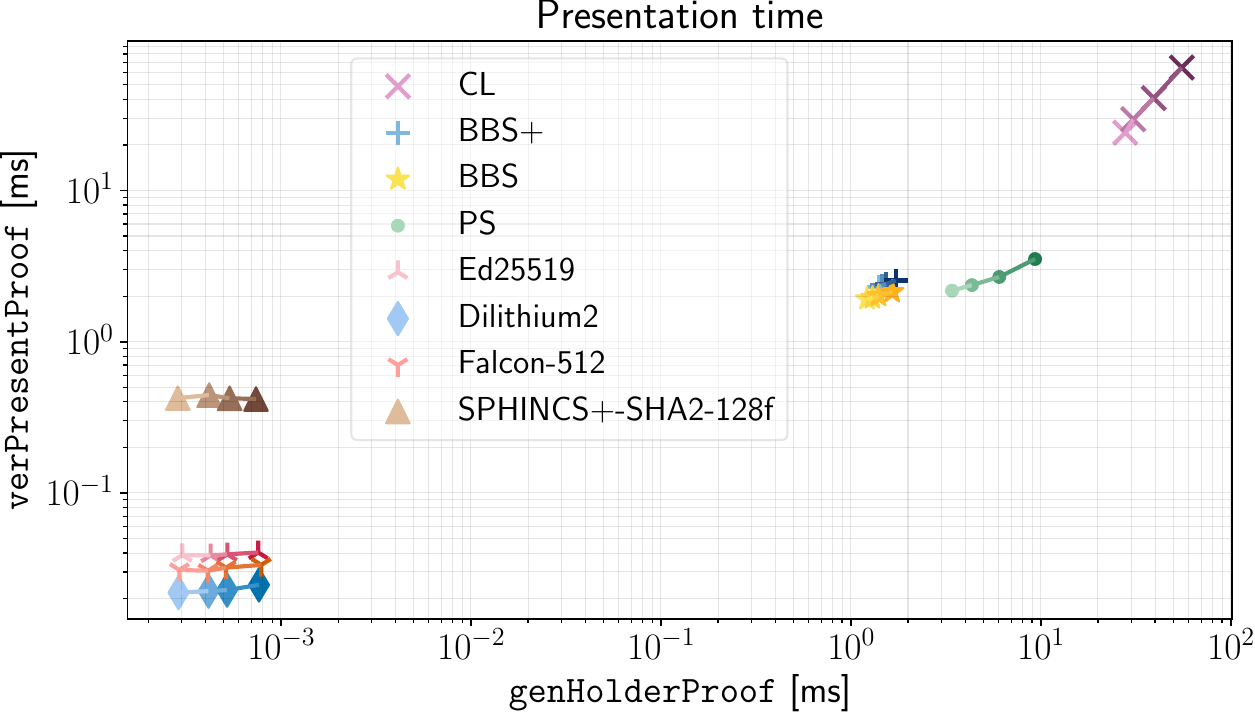}
	\caption{Scatter plot of Issuing and Presentation performance test results on Ryzen 7 5800X for all algorithms. Lower values are better (shorter run time), to the bottom left. Points are median values over all possible disclosed number of attributes in the range $n_D\in\{1,n_A\}$ with $n_A = [4,8,16,33]$. Darker colors correspond to higher $n_A$. \texttt{\merTree} algorithms are faster in both generation ($x$ axis) and verification ($y$ axis) of signatures and presentations. Quantum-Safe algorithms are very competitive with EdDSA, except \sphincs.}
	\label{fig:presentation_benchmark}
\end{figure}

We measure the speed of key generation, and signature 
and presentation proof generation and verification - see 
Figure~\ref{fig:presentation_benchmark}. 
The presentation phase is of particular interest since it is expected to occur frequently on constrained devices; we show results in
Table~\ref{tab:presentation_generation_arch_comparison}. 
There is approximately an order of magnitude difference in performance between a modern desktop CPU (Ryzen 7 5800X) and an ARM raspberry pi 4B, and another between the pi 4B and pi 3B+. $\texttt{\merTree}$ results are very close for each algorithm except \sphincs, which is reported separately; similarly \bbs~and \BBS~are very close, so only the former is reported.

\begin{table}[!b]
\caption{Presentation Proof generation and verification times by CPU with $n_A=33$. Times in ms are median over $n_D\leq n_A$. Falcon and Dilithium results are not significantly different from EdDSA; \BBS results are not significantly different from \bbs.}
\label{tab:presentation_generation_arch_comparison} \centering
    \begin{tabular}{lrrrr}
	\toprule
	\texttt{\cm: \genHolderProof}       & 5800X     &  pi 4B	    & pi 3B+   \\
	\midrule
	$\TT{\merTree}-EdDSA$    & 0.0007	   & 0.0044	& 0.0143 \\
        $\TT{\merTree}-\sphincs$    & 0.0007	   & 0.0044	& 0.0174 \\
	\bbs            & 1.6508	        & 11.2507	& 92.9144 \\
        PS              & 9.3314  	   & 34.6971	  & 303.4390 \\
        CL             & 55.270	   & 394.9524	& 1475.9164  \\
	\bottomrule
    \end{tabular}

\label{tab:presentation_verification_arch_comparison} 
    \begin{tabular}{lrrrr}
	\toprule
	\texttt{\cm: \verPresProof}        & 5800X     & pi 4B	    & pi 3B+   \\
	\midrule
	$\TT{\merTree}-EdDSA$    & 0.040	& 0.2911	& 1.0715 \\
        $\TT{\merTree}-\sphincs$    & 0.4176	   & 7.0385	& 19.3736 \\
	\bbs            & 2.1360	       & 15.7105	   & 117.4054 \\
        PS               & 3.5201        & 19.4249	& 210.9906 \\
        CL             & 64.969	   & 456.0057	& 1687.3662  \\
	\bottomrule
    \end{tabular}
\end{table}

The speed of hashing and of generating Merkle inclusion paths is negligible compared with generating and verifying the digital signature in \texttt{\comList} and \\
\texttt{\merTree}; we therefore only report the results of \texttt{\merTree} speed tests.


For \SDSig~over elliptic curves - BBS, BBS+, and PS - we also provide a comparison of the number of operations in Table~\ref{tab:ops_number}. Scalar multiplication ($M$) is $nQ$ for a curve point $Q$ and a scalar $n$, and multi-scalar multiplication of $n$ points, $MSM(n)$, is an operation designed to be more efficient than the sum of $(n-1)$ separate scalar multiplications.

The relation between the costs of the operations over the curve BLS12-381 are the following: $P > MSM > M >> A >>$ field operations, which we do not include. Roughly speaking, $P \approx 2~MSM(30)$, $MSM(30) \approx 3~M$, $M \approx 200~A$, $A \approx 100$ field operations, and operations over $\mathbb{G}_2$ cost approximately 2 to 3 times as operations over $\mathbb{G}_1$. These ratios consider multiplications $M$ and multi-scalar multiplications by random scalars.

\begin{table*}[!b]
\caption{Number of multiplication ($M$), multi-scalar multiplication ($MSM$), addition ($A$), pairing ($P$), and random sampling ($R$) operations over elliptic curves for \bbs, \BBS, and \PS, with number of disclosed ($n_D$), undisclosed ($n_U$), and total number of attributes ($n_A$). All operations are in the group $\mathbb{G}_1$ unless otherwise subscripted with $\mathbb{G}_2$.}\label{tab:ops_number}\centering
    \begin{tabular}{||p{1cm}|p{2cm}|p{2cm}|p{3cm}| p{3cm}||}
        \hline
        \texttt{\textbf{\SDSig}} & \texttt{\sign}           & \texttt{\ver}       & \texttt{\genHolderProof}       & \texttt{\verPresProof}   \\
        \hline
        $\TT{BBS}$                    & $MSM(n_A) + M + A$  & $MSM(n_A) + A + A_{\mathbb{G}_2} + M_{\mathbb{G}_2} + 2P$     & $MSM(n_A) + 3M + 2A + MSM(n_D) + MSM(2 + n_U)$  & $MSM(n_D) + MSM(3 + n_U) + 2A + 2P$    \\
        \hline
        $\TT{BBS+}  $                 & $MSM(1 + n_A) + M + A$     & $MSM(1 + n_A) + A + A_{\mathbb{G}_2} + M_{\mathbb{G}_2} + 2P$ & $MSM(1 + n_A) + 3M + 2A + MSM(2 + n_U) + MSM(2)$      & $MSM(n_D) + MSM(3 + n_U) + MSM(3) + 3A + 2P$ \\
        \hline
        $\TT{PS}$                     & $M_{\mathbb{G}_2} + R_{\mathbb{G}_2}$ & $MSM(n_A) + A + 2P$    & $3M_{\mathbb{G}_2} + A_{\mathbb{G}_2} + MSM(1 + n_U)$ & $MSM(2 + n_U) + A + MSM(n_D) + 2P$   \\
        \hline
    \end{tabular}
\end{table*}

\subsection{Presentation Size}
\label{subsec:size}

\begin{figure}[!ht]
    \centering
	\includegraphics[width=0.6\textwidth]{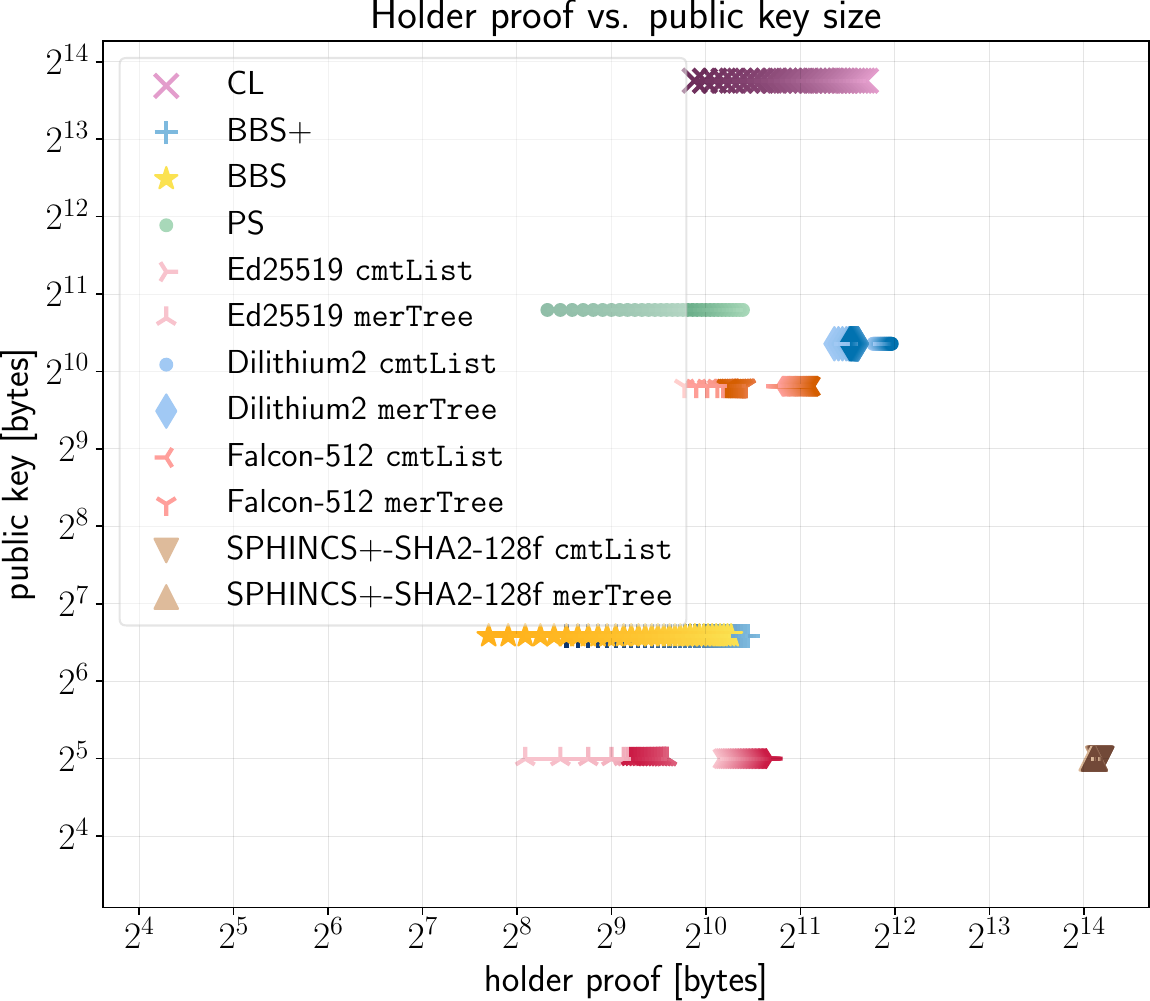}
	\caption{VP proof size - Eq. (\ref{eq:proof_size_cmtList}) to (\ref{eq:proof_size_PS}) - vs. public key size - Eq. (\ref{eq:pk_CL}), (\ref{eq:pk_BBS}), (\ref{eq:pk_BBS+}), (\ref{eq:pk_PS}) for \SDSig. Both are required by the Verifier, but the public key may be cached over several presentations, and the Holder-Verifier channel is more likely to have bandwidth constraints than the Issuer-Verifier channel in a digital wallet scenario, so a smaller proof size is more significant than a smaller public key size. Lighter hues are more disclosed attributes (higher $n_D$). \texttt{\SDSig} have smaller holder proofs for higher $n_D$ as fewer ZKP need to be generated for undisclosed attributes; \texttt{\comList} mechanisms follow the opposite trend, as more salts need to be disclosed. Common values used for comparison, in bytes: salt size ${s=16}$; digest size ${d=32}$; number of attributes ${n_A = 33}$.}
 
	\label{fig:selective-disclosure-proof-size}
\end{figure}

\paragraph{Presentation Proof.} We compare VP size contributions
for each \texttt{\cm}, with trends summarized in Figure~\ref{fig:selective-disclosure-proof-size}. 

Attribute size is arbitrary, so $\TT{\DA}$ is not included. For commitment based mechanisms, one disclosed salt $\TT{\DS}$ per disclosed attribute must be included; therefore, VP size tends to grow with $n_D$ for $\TT{\comList}$ and $\TT{\merTree}$, while it decreases for $\TT{\CL,\PS,BBS}$ and $\TT{\BBS}$ due to one zero-knowledge proof per undisclosed attribute.

Presentation proof size, for the cases in which multiple constructions are available, namely BBS and PS, is calculated according to the libraries we have tested. 

The presentation proof size for each cryptographic mechanism is computed as follows:

\begin{itemize}
    
    \item   $\TT{\comList}$: 
    one digest per attribute in the credential, a signature of the list of digests, one disclosed salt per disclosed attribute:
    \begin{align}\label{eq:proof_size_cmtList}
        \left|\TT{\SDS}\right|+\left|\sig \right| + \left| \TT{\DS}\right|=dn_A+|\sig|+sn_D
    \end{align}
    \item   $\TT{\merTree}$:
    one tree root (of digest size), a signature of the tree root, one disclosed salt per disclosed attribute, an inclusion proof for disclosed attributes. The size of an inclusion proof for a single attribute is equal to the tree height $\lceil\log_2(n_A)\rceil$ times the digest size; a simple implementation may return a separate proof per disclosed attribute, an optimized implementation will be more compact. An upper bound is therefore:
    \begin{align}
        \left|\TT{\SDS}\right|&+\left|\sig \right|+ \left|\TT{\DS}\right|+\left|P\right| = \nonumber \\ 
        &=d+|\sig|+d n_D+\lceil\log_2(n_A)\rceil dn_D \label{eq:proof_size_merTree_upper_bound}
    \end{align}
    \item $\TT{\CL}$: in order to make a fair comparison between the algorithms, we consider a modulus of $\left|n\right|= 3072$ bits to have a security level of 128 bits. Therefore, a NIZKP of knowledge of a signature and of the undisclosed attributes (Eq.~(\ref{eq:genHolderProof_cl})) is given by: 
    \begin{itemize}
        \item a digest $c\in \{0,1\}^{256}$ (32 bytes);
        \item an element $A'\in \mathbb{Z}_n$ (384 bytes), an element $\widehat{e}\in \{0,1\}^{457}$ (58 bytes), and $\widehat{v}'\in \{0,1\}^{3744}$ (468 bytes) 
        \item an element $\widehat{a}_i \in \{0,1\}^{593}$ (75 bytes) for each undisclosed attribute.
    \end{itemize}
Therefore, the presentation proof size is, in bytes: 
    \begin{align}\label{eq:proof_size_CL}
        \left|c\right|+\left|A'\right| + \left|\widehat{e}\right|+ \left|\widehat{v}'\right|+ \left|\widehat{a}_i\right|(n_A-n_D)= \nonumber \\ =32+384+58+468+75(n_A-n_D).
    \end{align}
    \item $\TT{BBS:}$ a NIZKP of knowledge of a signature and of the undisclosed values (Eq.~(\ref{eq:genHolderProof_bbs})) is given by:
    \begin{itemize}
        \item three elements $\overline{A}, \overline{B}, T\in\mathbb{G}_1$;
        \item two elements $\widehat{r},\widehat{e}\in \mathbb{Z}_p$;
        \item one $\widehat{a}_i\in \mathbb{Z}_p$ for each undisclosed attribute.
    \end{itemize}
    $\TT{BBS}$ can be implemented using the pairing-friendly elliptic curve BLS12-381, with the prime order of the subgroup of $\mathbb{G}_1$ being $p\in \{0,1\}^{256}$. Therefore, the elements in $\mathbb{G}_1$ - i.e., $\overline{A},\overline{B}$ and $T$ - can be represented as 48-byte strings and the integer elements as 32-byte strings. Therefore, the presentation proof size is, in bytes:
    \begin{align}\label{eq:proof_size_BBS}
        \left|\overline{A}\right| + \left|\overline{B}\right| + \left|T\right|+ \left|\widehat{r}\right|+ \left|\widehat{e}'\right|+ \left|\widehat{a}_i\right|(n_A-n_D)= \nonumber \\ =3\cdot 48+32(2+n_A-n_D).
    \end{align}
    \item   $\TT{\BBS}$: a NIZKP of knowledge of a signature and of the undisclosed values (Eq.~(\ref{eq:genHolderProof_bbs+})) is given by:
    \begin{itemize}
        \item five elements $A',\overline{A},d,T_1,T_2\in\mathbb{G}_1$;
        \item four elements $\widehat{e},\widehat{r}_2,\widehat{r}_3, \widehat{s}' \in \mathbb{Z}_p$;
        \item one $\widehat{a}_i\in \mathbb{Z}_p$ for each undisclosed attribute.
    \end{itemize}
    As with $\TT{BBS}$, $\TT{\BBS}$ can be implemented using the pairing-friendly elliptic curve BLS12-381, with the prime order of the subgroup of $\mathbb{G}_1$ being $p\in \{0,1\}^{256}$. Therefore, the elements in $\mathbb{G}_1$ - i.e., $A',\overline{A},d$ - can be represented as 48-byte strings and the integer elements as 32-byte strings.
    Therefore, the presentation proof size is, in bytes:
    \begin{align}\label{eq:proof_size_BBS+}
        \left|A'\right| +
        \left|\overline{A}\right| + \left|d\right| + \left|T_1\right|+ \left|T_2\right| + \left|\widehat{e}\right|+
        \left|\widehat{r}_2\right|+ \left|\widehat{r}_3\right|+ \nonumber \\ 
        \left|\widehat{s}'\right|+
         \left|\widehat{a}_i\right|(n_A-n_D)= 5\cdot 48 +32(4+n_A-n_D).
    \end{align}
    \item $\TT{\PS}$: a NIZKP of knowledge of a signature and of the undisclosed values
    (Eq.~(\ref{eq:genHolderProof_ps})) is given by:
    \begin{itemize}
        \item two elements $\sigma_1', \sigma_2'\in\mathbb{G}_2$;
        \item two elements $J,T\in\mathbb{G}_1$;
        \item an element $\widehat{t}\in \mathbb{Z}_p$;
        \item one $\widehat{a}_i\in \mathbb{Z}_p$ for each undisclosed attribute.
    \end{itemize}
    Also PS can be implemented using the pairing-friendly elliptic curve BLS12-381 with the prime order of the subgroup of $\mathbb{G}_1$ being $p\in \{0,1\}^{256}$. Therefore, the elements in $\mathbb{G}_2$ - i.e., $\sigma_1',\sigma_2'$ - can be represented as 96-byte strings, the elements in $\mathbb{G}_1$ - i.e., $J,T$ - can be represented as 48-byte strings and the integer elements as 32-byte strings. 
    Therefore, the presentation proof size is, in bytes:
    \begin{align}\label{eq:proof_size_PS}
        \left|\sigma_1'\right| + \left|\sigma_2'\right| + \left|J\right|+ \left|T\right|+ \left|\widehat{t}\right|+ \left|\widehat{a}_i\right|(n_A-n_D)= \nonumber \\ =2\cdot 96+2\cdot 48+32(1+n_A-n_D).
    \end{align}
\end{itemize}

\paragraph{Issuer Public keys $(\TT{pk_{Iss}})$.}A VP header may contain either $\TT{pk_{Iss}}$ or a reference to it. For instance, a JWS\\
\cite{RFC_7515} header may contain $\mathtt{pk_{Iss}}$ as JWK or X.509 certificate, or a url as JKU, or a certificate thumbprint, etc. Note that in standardized digital signatures, public parameters are assumed to be known and common to all issuers and verifiers, typically written in source code of implementing libraries, and need not be fetched repeatedly. For instance, the Digital Signature Standard~\cite{FIPS_186-5} (DSS) specifies methods for signature generation and verification, while specifications for the generation of domain parameters e.g., for ECDSA and EdDSA are included in SP 800-186~\cite{NIST_SP_800-186}. We do not describe these algorithms in detail ~\cite{FIPS_186-5,NIST_SP_800-186}.  

$\TT{pk_{Iss}}$ size may be calculated as follows.
\begin{itemize}
    \item $\TT{\merTree, \comList}$: the public key of an EdDSA digital signature is a 32-byte point of the curve ed25519. A \sphincs-128~\cite{bernstein2019sphincs+} public key is 32 bytes: two 128-bit numbers, a seed and a tree root. A Dilithium2~\cite{lyubashevsky2020crystals} public key is 1312 bytes: a 32-byte seed and a vector of polynomials. A Falcon-512~\cite{fouque2018falcon} public key is 897 bytes: 512 14-bit integers (polynomial coefficients), and 1 header byte.
    \item $\TT{\CL}$: A CL public key (Eq. (\ref{eq:pk_CL})) is $(n,R_1,\dots,R_{m},S,Z)\in \TT{\mathbb{Z}_n^{m+3}}$ of size $384(m+3)$ bytes.
    \item $\TT{BBS}$: A BBS public key (Eq. (\ref{eq:pk_BBS})) is $w=g_2^x\in \mathbb{G}_2$ of size $96$ bytes, where $\mathbb{G}_2$ is obtained using curve BLS12-381.
    \item $\TT{\BBS}$: A BBS+ public key (Eq. (\ref{eq:pk_BBS+})) is $w=g_2^x$  of size $96$ bytes, where $\mathbb{G}_2$ is obtained using curve BLS12-381.
    \item $\TT{\PS}$: A PS public key (Eq. (\ref{eq:pk_PS})) is  $(X,Y_1,\dots,Y_m)\in \mathbb{G}_1^{m+1}$ of size $48(m+2)=96+48m$  bytes if PS is instantiated using curve BLS12-381.
    \end{itemize}

\subsection{Trade-offs}
\label{subsec:tradeoffs}
\paragraph{Switching $\mathbb{G}_1$ and $\mathbb{G}_2$.} 

When considering the groups derived from the elliptic curve BLS12-381, which is used in the implementations we have tested, performing computations over the group $\mathbb{G}_1$ is more efficient than performing the same computations in $\mathbb{G}_2$. Also the size of elements in $\mathbb{G}_1$ (48 bytes) is smaller than the size of elements in $\mathbb{G}_2$ (96 bytes). However, the bilinearity of the pairing operation allows to define the BBS, BBS+ and PS signature algorithms (and the associated NIZKP) inverting the roles of $\mathbb{G}_1$ and $\mathbb{G}_2$, and this leads to the identification of some trade-offs that it is worth to consider when instantiating these schemes.

Note that if PS is implemented inverting the role of $\mathbb{G}_1$ and $\mathbb{G}_2$, then the signatures are $(\sigma_1,\sigma_2)\in \mathbb{G}_1^2$ and the public keys are $(X,Y_1,\dots,Y_m)\in \mathbb{G}_2^{m+1}$. Then the size of the signature is reduced to 96 bytes, the public key size is increased to $96+96m$ bytes and the presentation proof dimension is unchanged. Since the computations are more efficient when computed in $\mathbb{G}_1$, the signature generation (VC issuance) will be faster, while the presentation proof generation will be slower.

For what concerns BBS and BBS+ signatures, switching the two groups $\mathbb{G}_1$ and $\mathbb{G}_2$ reduces the size of the public key to 48 bytes, increases the size of both the signature and the presentation proofs, and slows down the computations for the generation of both the signatures and the NIZKPs, since they would be performed in $\mathbb{G}_2$.

\paragraph{Structure of NIZKP.}
In Section \ref{sec:sd-signatures} we describe the structure of the NIZKPs corresponding to each selective disclosures signature. As it is possible to note from Equation (\ref{eq:genHolderProof_cl}), the NIZKP of CL has an element $c$ in $\mathbb{Z}_p$, while the NIZKP of BBS (Eq. (\ref{eq:genHolderProof_bbs})) and PS (Eq. (\ref{eq:genHolderProof_ps})) have an element $T$ in $\mathbb{G}_1$ and BBS+ has two elements $T_1,T_2\in\mathbb{G}_1$ (Eq. (\ref{eq:genHolderProof_bbs+})).
\footnote{The NIZKP that we have described in Equation \ref{eq:genHolderProof_bbs+} is obtained by proving two relations as in Figure \ref{fig:sigma_fiat_shamir}. Therefore the prover must generate two commitments $T_1,T_2$, one for each relation, but it can use the same challenge $c=\HH(\pp,T_1,T_2)$.}
The NIZKP derived from the sigma protocols for linear relations described in \ref{sigmaFS} (see Figure \ref{fig:sigma_fiat_shamir}) instructs the prover to compute a random commitment $T\in \mathbb{G}$, derive deterministically a challenge $c=\HH(\pp,T)\in \mathbb{Z}_p$, where $\pp$ are public parameters known both to the prover and to the verifier, and from these compute the response $(r_1,\dots,r_n)\in\mathbb{Z}_p^n$. 
Finally the prover can build its proof $\pi$ in two equivalently secure ways:
\begin{enumerate}
    \item if the prover sends $T,(r_1,\dots,r_n)$, the verifier can compute $c$ and verify the validity of the proof $\pi$;
    \item if the prover sends $c,(r_1,\dots,r_n)$, the verifier can retrieve $T$ and verify the validity of the proof $\pi$. Note that this does not require inverting $\HH$ - see Figure \ref{fig:sigma_fiat_shamir}.
\end{enumerate}
When the representation of an element of the group $\mathbb{G}$ is bigger in size than an element of $\mathbb{Z}_p$, by choosing the second approach the proof is smaller.

However, it might be preferable to choose the first approach in a context in which $(i)$ the Holders can present multiple VCs all at once (by proving multiple statements), for example to prove predicates which relate different VCs, and $(ii)$ the Verifiers, in case of failure, want to identify the statements whose proof was incorrect\footnote{Private communication with one of the developers of the Docknetwork Library.}. This approach improves error checking.

In this context, a single challenge $c=\HH(\pp,T1,\dots,T_n)$ is generated according to the commitments associated to each of the $n$ relations to be proved. Then, according to $c,T_1,\dots,T_n$, the Holder generates the responses included in the VP. If the Verifier receives the commitments $T_1,\dots, T_n$ and the responses, it can compute the challenge and verify the correctness of each proof individually, identifying which proofs have failed, if any. 
Instead, if we use the second approach by sending to the Verifier the challenge $c$ and the responses, if any of the relations have not been proven correctly, it would be impossible to identify which statements caused the failure. From the responses and the challenge $c$, the Verifier can reconstruct the commitments $T_1,\dots,T_n$ obtaining $\HH(\pp,T_1,\dots,T_n)=c'\ne c$; this means that one or more statements have not been proven, but the Verifier can not identify which ones.

\subsection{Assessment Summary}
\label{subsec:assessment_summary}
We find that \texttt{\comList} and \texttt{\merTree} are very fast to compute, cryptographically agile and with quantum-safe options, easier to implement than \texttt{\SDSig} but more cumbersome for the Issuer to manage. \texttt{\merTree} is reliably smaller in size, but widely adopted in standards and RFCs. Predicates must be defined by the issuer, and unlinkability requires the issuer to provide a supply of single-use credentials with new attribute salts and signatures in advance. While \texttt{\CL} is particularly computationally expensive and large in size, \texttt{\bbs} is computationally feasible and compact; in both cases, predicate proofs can be provided by the holder, and the randomness for unlinkability is also generated by the holder based on a single selective disclosure signature. Our assessment is summarized qualitatively in Table~\ref{tab:summary_analysis}.

\begin{table}[!b]

\caption{\texttt{\cm} assessment summary. }
\label{tab:summary_analysis} \centering
    \begin{tabular}{lccccc}
	\toprule
	Feature          & \texttt{\comList}    &   \texttt{\merTree}   & \texttt{\CL}	    & \texttt{\bbs(+)}	    & \texttt{\PS}	    \\
	\midrule
    Standard        &  \cellhigh    & \cellmeh     & \celllow    & \cellmeh    & \celllow   \\
    Agile         &  \cellhighhh    & \cellhighhh     & \cellloww    & \cellhigh    & \cellhigh  \\
    Unlinkable   &  \cellmeh    & \cellmeh     & \cellhigh    & \cellhigh    & \cellhigh \\
    Predicates      &  \cellmeh    & \cellmeh 	& \cellhigh    & \cellhigh    & \cellhigh \\
    Fast           &  \cellhighhh    & \cellhighhh 	& \celllow   & \cellmeh    & \cellmeh   \\
    Compact         &  \celllow    & \cellhigh     & \celllow    & \cellhigh    & \cellhigh  \\
    Quantum-safe    &  \cellhigh    & \cellhigh     & \celllow    & \celllow    & \celllow  \\
	\bottomrule
    \end{tabular}
\end{table}

\section{Conclusion}
\label{sec:conclusion}
As the digital landscape continues to expand, the significance of an individual's digital identity cannot be overstated, particularly in the realms of e-government and e-commerce. The emergence of digital identity wallets represents a pivotal shift in identity management, enabling data subjects to exercise control over the information they disclose in a secure and privacy-preserving manner. This paradigm shift is exemplified by the eIDAS2 regulation, which proposes the EUDI wallet to enhance cross-border interoperability. Thus, there is a need for service providers to strike a balance between protocol sophistication, implementation intricacy, and resource constraints. The ARF document published by the EU Commission underscores the importance of cryptographic mechanisms that enable selective disclosure of verifiable credentials, which is a crucial component of providing privacy-preserving solutions.

To fill this gap, this paper provides an overview of cryptographic mechanisms to enable selective disclosure of verifiable credentials, which serve as digital counterparts to physical credentials and are safeguarded by cryptographic techniques. We analyzed a total of six mechanisms: hash-based hiding commitments $\TT{\comList}$, $\TT{merTree}$, as well as CL, BBS/BBS+, and PS signatures. For each mechanism, we defined the credential and presentation structures, and summarized the operations to be performed to issue VCs and provide VPs. In order to assist stakeholders with the knowledge needed to make informed decisions regarding the selection and implementation of cryptographic mechanisms based on their specific use cases and system requirements, we compared the cryptographic mechanisms w.r.t. several features such as standardization, cryptographic agility, performance, and quantum safety.

In summary, our solution analysis indicates that $\TT{\comList}$ and $\TT{merTree}$ are highly efficient in terms of computation speed and offer cryptographic agility, along with quantum-safe options. These mechanisms are relatively straightforward to implement, although they may pose greater management challenges for the Issuer. On the other hand, CL exhibits high computational expense and size, whereas BBS presents a more feasible and compact alternative. In both cases, the holder has the capability to provide predicate proofs, and randomness for achieving unlinkability is generated by the holder based on a single selective disclosure signature. 

\paragraph{Acknowledgements}
The first author acknowledges support from Eustema S.p.A. through the PhD scholarship and is a member of GNSAGA of INdAM.

This work has been partially supported by “Futuro \& Conoscenza Srl”, jointly created by the FBK and the Italian Government Printing Office and Mint, Italy.

This work was partially supported by project SERICS (PE00000014) under the MUR National Recovery and Resilience Plan funded by the European Union - NextGenerationEU.

The authors would like to thank Vasilis Kalos and Lovesh Harchandani for the insightful discussions on BBS specification and implementations.

\appendix
\section{NIZKP from Sigma Protocols via Fiat-Shamir transform}
\label{sigmaFS}
The NIZKP that we use in this paper are derived from three-step interactive protocols called \emph{sigma protocols}. Sigma protocols allow a prover to prove a statement to a verifier based on the knowledge of a secret (e.g. ``given $h,g\in \mathbb{G}$, I know $x\in\mathbb{Z}_p$ such that $h=g^x$''). In order to do that, it sends a commitment $T$ to the verifier who returns a random challenge $c$. Finally, according to $T$, $c$ and the statement to be proven, the prover sends a response $r$ to the verifier who checks the validity of the transcript $(T,c,r)$ for that specific statement and accept or rejects the interactive proof. 

A sigma protocol can be turned into a non-interactive protocol by applying the \emph{Fiat-Shamir transform}. Informally, the Fiat-Shamir transform prescribes to replace the generation of the challenge by the verifier with a computation of a digest of $T$ and other public data via a cryptographic hash function $\HH$ computed by the prover. Below we describe more in detail what is a sigma protocol and what is the Fiat-Shamir transform.

\paragraph{Sigma protocols.}
A sigma protocol ${(P,V)}$ for the relation $\mathcal{R}$ is defined by
\begin{itemize}
    \item the relation ${\mathcal{R}\subset W\times Y}$, where ${Y}$ is called the \emph{set of statements} and ${W}$ the \emph{set of witnesses};
    \item two algorithms describing the behaviour of the actors involved: the prover $P$ and the verifier $V$.
\end{itemize} 
We say that ${(w,y)\in \mathcal{R}}$ if and only if ${w}$ is a \emph{witness} for the \emph{statement} ${y}$. A sigma protocol for a relation ${\mathcal{R}}$ allows the prover to convince a verifier about the knowledge of a secret witness ${w}$ for a public statement ${y}$. 

The sigma protocols are three steps protocols with the following structure: the prover ${P}$ computes a message $T$ called  \emph{commitment} and sends it to the verifier ${V}$. Once ${V}$ has received the commitment, it chooses a random \emph{challenge} ${c}$ and sends it to the prover ${P}$. Then, ${P}$ computes the response ${r}$ and sends it to ${V}$. Finally ${V}$ outputs 1 (accept) or 0 (reject) which must be computed according to statement and the \emph{transcript} ${(T,c,r)}$ generated by the interaction. A secure sigma protocol must satisfy the following properties:
\begin{itemize}
    \item \emph{completeness}: when a prover knows a witness ${w}$ for a statement ${y}$, the verifier will output $\TT{accept}$ at the end of the protocol.
    \item \emph{knowledge soundness}: if the verifier outputs $\TT{accept}$, it is assured that the prover actually knows a witness $w$ for the public statement $y$;
    \item \emph{honest-verifier zero-knowledge}: the interaction with an honest verifier in a sigma protocol execution does not leak any information about the witness known by the prover.
\end{itemize}

An example of secure sigma protocol, which is the building block of the NIZKPs mentioned in Section \ref{sec:sd-signatures}, is the \emph{sigma protocol for linear relations}~\cite{boneh2020graduate}, a generalisation of the well known Schnorr sigma protocol~\cite{Schnorr_91}. 

\begin{figure*}[tp]
\input{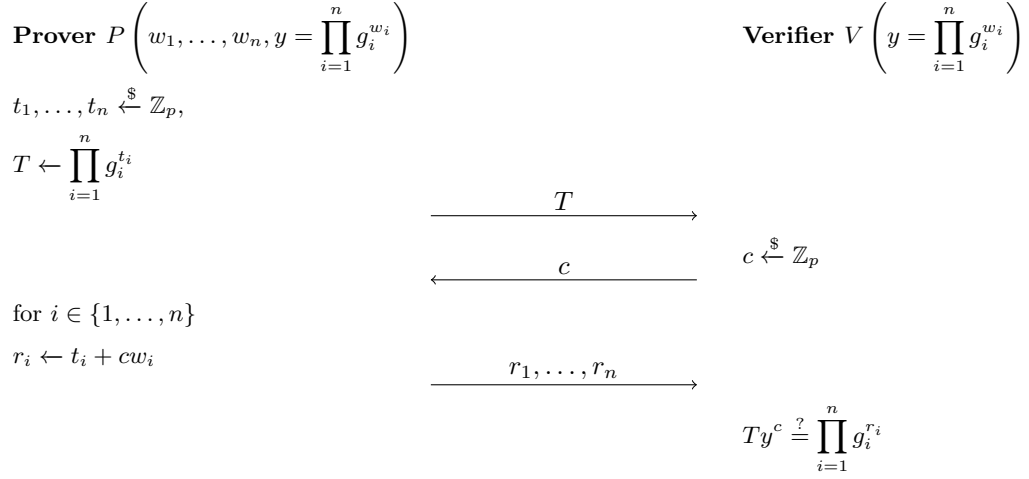}
\caption{Sigma protocol for linear relations. The Simulator used to prove the zero-knowledge property is defined as follows: it generates uniformly at random $s_1,\dots,s_n,c\in\mathbb{Z}_p$ and sets $T=y^{-c}\prod_{i=1}^n g_i^{s_i}$. The transcript $(T,c,s_1,\dots,s_n)$ is indistinguishable from a real transcript since $c$ is random and $T$ is random as well since it is univocally determined by $s_1,\dots,s_n$ which are chosen uniformly at random. The transcripts verify and have been created without knowing $w_1,\dots,w_n$.}
\label{fig:LinRel_sigma_protocol}

\end{figure*}

Given $n\in \mathbb{N}$ and $g_1,\dots,g_n\in \mathbb{G}$, where $\mathbb{G}$ is a group of order $p$, $g_i \in \mathbb{G}$ are public parameters, the sigma protocol for linear relations is defined by $\mathcal{R}=\{((w_1,\dots,w_n),y)\in \mathbb{Z}_p^n\times \mathbb{G} \mid y=\prod_{i=1}^n g_i^{w_i}\}$ and by the algorithms $(P,V)$ presented in Figure \ref{fig:LinRel_sigma_protocol}.

\paragraph{Fiat-Shamir transform.}

In 1986 A. Fiat and A. Shamir introduced in \cite{fiat1986prove} a technique 
to convert \emph{identification schemes}, used to identify a user according to a secret only she knows\footnote{From sigma protocols is possible to derive identification schemes. In identification schemes, the statements of the sigma protocols are the public keys of users, and the user who knows the witness for such statement proves its identity.}, into digital signature schemes. However, the same technique can be applied to secure sigma protocols - satisfying the completeness, knowledge soundness, and HVZK properties above - to obtain non-interactive zero-knowledge proofs \cite{boneh2020graduate}.

The Fiat-Shamir transform 
substitutes the verifier with a random oracle during the second step of the sigma protocol, 
in which the verifier generates a random challenge; this operation can be performed by a random oracle, considered as a trusted third party that can be impersonated by a cryptographic hash function. The prover, instead of sending the commitment to the verifier, computes a cryptographic hash of the commitment, together with other public data $\TT{pp}$ that identify the protocol execution; the output becomes the sigma protocol challenge. The prover then computes the response to the challenge and sends the transcript to the verifier, who can compute using the same hash function the same random challenge and verify that the prover does know a witness.

\begin{figure*}[tp]
\input{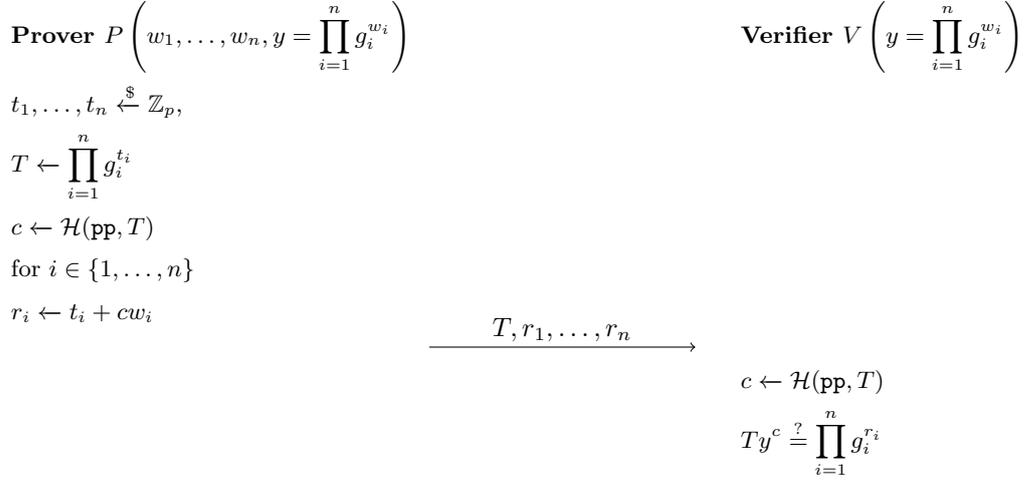}
\caption{NIZKP for linear relations. The term $\TT{pp}$ is given by the public parameters such as $g_1,\dots,g_n,y$.\\
It would be possible for the prover to create a NIZK proof by sending $(c,r_1,\dots,r_n)$. In this case the verifier must compute $T\xleftarrow{}\prod_{i=1}^n g_i^{r_i}y^{-c}$ and check that $c\stackrel{?}{=}\HH(\TT{pp},T)$.
}
\label{fig:sigma_fiat_shamir}
\end{figure*}

Applying the Fiat-Shamir transform to the secure sigma protocol in Figure \ref{fig:LinRel_sigma_protocol} (i.e. which satisfies the completeness, knowledge soundness and honest-verifier zero-knowledge properties) yields the NIZKP in Figure~\ref{fig:sigma_fiat_shamir}, secure according to the threat model described in Section \ref{subsec:NIZKP}.

Below we provide a more formal description of what it means for a sigma protocol to be HVZK compared to the one given above. 

\paragraph{Honest-verifier zero-knowledge.} The term HVZK refers to protocols in which a prover proves to an honest verifier, i.e. a verifier that follows the protocol instructions, the knowledge of some secret information $w$ without disclosing any other information. 
This concept is formalised starting from a very reasonable observation: the only way an eavesdropper, who observes their interaction, can try to extract some information about $w$ is by observing real executions of the protocol performed by the prover and the verifier. Then, based on the information exchanged between the two parties (a transcript of the protocol), the eavesdropper tries to learn something about the secret known by the prover. 

However, if there exists an efficient algorithm, referred to as \emph{Simulator}, which does not take $w$ in input and is capable to create transcripts that are indistinguishable from the transcripts of real protocol executions, then we say that the protocol is \emph{honest-verifier zero-knowledge}. The reason is the following: observing the prover interacting with the verifier, i.e. a real transcript, is indistinguishable from a transcript generated by the Simulator, i.e. a simulated transcript. This means that it is possible to extract the same amount of information from the two. However, the simulator does not know the secret $w$, therefore it is impossible, based on the transcripts it generates, to learn some information about $w$. This means that, for an eavesdropper, eavesdropping the conversation between the prover and the verifier and generating transcripts on its own by executing the Simulator on her laptop, gives her the same advantage in learning information about $w$. 

It is not trivial to determine whether a protocol satisfies the HVZK property, 
or indeed whether it even admits a simulator or not. 
The algorithm must be capable of producing transcripts indistinguishable from the ones generated in real protocol executions. 

An example of simulator for the sigma protocol for linear relations is provided in the caption of Figure \ref{fig:LinRel_sigma_protocol}.

\bibliographystyle{alpha}
\bibliography{Bibliography}

\end{document}